\documentclass[spanish,reqno]{amsart}
\usepackage[T1]{fontenc}
\usepackage[latin1]{inputenc}
\usepackage{amsmath}
\usepackage{amssymb}
\usepackage{amsfonts}
\usepackage{graphicx}
\usepackage{hyperref}
\usepackage{babel}

\begin{document}

\title{Introducción a la Supersimetr\'{\i}a}
\author{José A. Vallejo}
\address{Facultad de Ciencias\\
 Universidad Autónoma de San Luis Potos\'{\i} (México)\\
 Lateral Av. Salvador Nava s/n, SLP 78290.}
\email{jvallejo@fc.uaslp.mx}
\thanks{Parcialmente financiado por: Proyecto SEP-CONACyT Ciencia Básica (J2) 
2007-1 código 78791 (México) y Proyecto MTM2005-04947 Ministerio de Educación y
Ciencia (España). Este trabajo se encuentra en su versión final y no será 
sometido a otra publicación.}
\keywords{Superálgebra de Heisenberg, superespacios vectoriales, 
Mecánica Cuántica supersimétrica.}
\subjclass[2000]{81Q60,17B70,17B81,16W50.}
\begin{abstract}
Estas notas pretenden ser una introducción elemental al tema de la
supersimetr\'{\i}a adecuada a la formación de matemáticos, partiendo
de los conceptos básicos de la Mecánica Cuántica. El objetivo 
fundamental es obtener una realización de la superálgebra de
Heisenberg como una subálgebra del álgebra graduada de endomorfismos de 
un cierto superespacio vectorial (un ejemplo del teorema de Ado en
superálgebras de Lie), utilizando para ello el modelo de Witten
de la Mecánica Cuántica supersimétrica.\\
Las notas tienen su origen en un seminario para estudiantes del posgrado 
en Ciencias Aplicadas de la Facultad de Ciencias de la Universidad Autónoma
de San Luis Potosí impartido por el autor.
\end{abstract}
\maketitle

\setcounter{tocdepth}{1}
\tableofcontents{}

\newpage
\section{Introducción}

En los últimos tiempos, el tema de la supersimetr\'{\i}a (en general,
de todas las teor\'{\i}as {}``súper'' ) ha pasado por todos los
estados de reconocimiento posibles: desde las primeras expectativas,
quizás un tanto optimistas, a un parcial abandono o
a una inesperada resurrección. El hecho es que se han mantenido a
flote por dos razones fundamentales: una, la más mencionada, es la
belleza y consistencia del formalismo matemático que se emplea en
su descripción, pero la otra, mucho más importante, es su aplicación
a la explicación de fenómenos que quedan fuera del alcance de las
teor\'{\i}as clásicas. No se debe olvidar que ésta fue la motivación
original para su introducción, las teor\'{\i}as de supersimetr\'{\i}a
tienen una vocación eminentemente práctica.

Sin embargo, bien es cierto que es dif\'{\i}cil apreciar estas posibles
aplicaciones cuando se estudia alguno de los textos clásicos sobre
el tema (ver \cite{Kos 77}, \cite{GSW 87}, \cite{Wes-Bag 92}),
y mucho menos aún es fácil darse cuenta de lo natural que resulta
el punto de vista supersimétrico. Habitualmente, o bien se comienza
discutiendo las propiedades de las posibles extensiones del grupo
de Poincaré para contemplar las simetr\'{\i}as internas, o bien se
estudian las colisiones entre part\'{\i}culas a altas energ\'{\i}as
para ilustrar la necesidad de una unificación de las interacciones
en ese régimen. Pero las ra\'{\i}ces de la teor\'{\i}a son mucho más
simples, y se remontan a los intentos de F. Berezin de dar una tratamiento
unificado para los bosones y fermiones en el contexto de la Mecánica
Cuántica (v\'ease \cite{Ber 66} y aplicaciones dentro de la mec\'anica cl\'asica en
\cite{Cas 76}).

Precisamente, estas notas tienen como objetivo el presentar a un público matemático
las ideas más básicas de la supersimetr\'{\i}a en un contexto elemental, accesible
con sólo la base de unas nociones elementales de Álgebra y Análisis
Funcional. No es imprescindible tener conocimientos previos de F\'{\i}sica
Cuántica (basta un curso de Física Básica), los conceptos necesarios
para la comprensión de la terminolog\'{\i}a habitual (bosones, fermiones,
etc...) se irán introduciendo a medida que se necesiten. De este modo,
el autor espera que las notas sean útiles a los matemáticos que estén
interesados en el tema y que se aproximen a él por primera vez.

Con el fin de mantener un nivel asequible, sólo discutiremos la Mecánica
Cuántica Supersimétrica (SUSY QM), un estudio de las teor\'{\i}as
de campo requiere de un formalismo matemático más avanzado y un tratamiento
mucho más extenso.

Comenzaremos con un breve repaso de cómo se introdujo, históricamente, el espín.
Un lector que no esté interesado en los aspectos físicos del problema, puede omitir
su lectura sin problema; conceptos como el de operador de espín serán introducidos
más adelante dentro de un contexto puramente matemático. La primera parte de las notas
(secciones 1 a 9) contienen una descripción del formalismo básico de la Mecánica Cuántica
expresado de una manera más formal que la acostumbrada en los libros de texto de 
Física, con el fin de que un matemático sin formación previa en Física pueda comprender
el lenguaje de la segunda parte (el resto de secciones), donde se presenta la 
construcción de una realización de la superálgebra de Heisenberg como una subálgebra
de los endomorfismos graduados de un cierto superespacio vectorial, dando así
un ejemplo de la validez del Teorema de Ado en la categoría de superálgebras de Lie. La
sección \ref{C11} es en realidad una excusa para presentar los rudimentos de la teoría
del álgebra lineal en superespacios vectoriales. Un lector con conocimientos previos
del formalismo estándar de la Mecánica Cuántica puede pasar directamente a la sección
\ref{C11}.
La sección 10 tiene un carácter más especializado que el resto, su intención es
la de presentar la idea de supersimetría en un contexto mucho más amplio que
el de la Mecánica Cuántica (el de las teorías gauge) y su lectura puede
omitirse sin problemas en caso de que se desconozcan los conceptos que en
ella aparecen.

La última sección tiene por objeto mostrar como esta construcción no es un mero ejercicio
matemático, sino que tiene profundas consecuencias en Física (que van mucho más allá
de las limitadas aplicaciones que veremos).

Con el fin de que estas notas también puedan resultar de utilidad a aquellos físicos
interesados en conocer la base formal de los trabajos sobre SUSY QM, 
se ha evitado la presentación habitual en Matemáticas (definiciones, 
lemas, teoremas, etc.) que tiende a ahuyentar a este tipo de lector. El autor 
espera que esta decisión no implique que quien se ahuyente sea el lector de inclinaciones
matemáticas.

\subsection*{Agradecimientos}
Durante la elaboración de estas notas, he tenido el beneficio de numerosas
discusiones con Juan Monterde, Gil Salgado, Adolfo Sánchez Valenzuela y Jesús Urías. 
A todos ellos quisiera darles las gracias por sus acertados comentarios, sugerencias y
útiles críticas. De manera muy especial quisiera agradecerle al \'arbitro an\'onimo sus valiosos comentarios acerca de
la bibliograf\'ia, gracias a los cuales el lector puede disponer de una acertada selecci\'on de 
trabajos para profundizar en su concimiento de la SUSY QM y otros temas relacionados.\\
Por supuesto, cualquier incorrección, omisión o malinterpretaci\'on en el texto, es 
responsabilidad exclusiva del autor.

\section{El experimento de Stern-Gerlach y el espín\label{SG}}

En el año 1922 eran frecuentes en Física los experimentos
de interacción entre part\'{\i}culas cargadas y campos magnéticos;
por ejemplo, para una part\'{\i}cula cargada que pasa entre dos imanes
con polaridades opuestas, dispuestos verticalmente, la teor\'{\i}a 
electromagnética clásica,
basada en el hecho de que la part\'{\i}cula no es puntual sino que
se describe por una cierta distribución de carga, predice lo siguiente:

\begin{enumerate}
\item Si la part\'{\i}cula no gira sobre si misma, no hay desviación vertical. 
\item Si la part\'{\i}cula tiene un movimiento dextrógiro, se desv\'{\i}a
hacia arriba. 
\item Si la part\'{\i}cula tiene un movimiento levógiro, se desv\'{\i}a
hacia abajo. 
\end{enumerate}
La magnitud de la desviación depende de la distribución de carga de
la part\'{\i}cula y de la velocidad angular de giro. Veámoslo con
más detalle: la teor\'{\i}a clásica supone que cada part\'{\i}cula
tiene un momento magnético permanente $\boldsymbol{\mu}$, dado por
$\boldsymbol{\mu}=\gamma\mathbf{S}$ donde $\mathbf{S}$ es el momento
angular intr\'{\i}nseco debido al giro de la part\'{\i}cula %
\footnote{Denotaremos los vectores tridimensionales en negrita.%
}. A $\gamma$ se le denomina coeficiente giromagnético o factor de
Landé. Cuando la part\'{\i}cula se coloca en el seno de un campo magnético
externo $\mathbf{B}$ el trabajo total que se efectúa es\[
W=-\boldsymbol{\mu}\cdot\mathbf{B},\]
 el momento de torsión viene dado por\[
\boldsymbol{\tau}=\boldsymbol{\mu}\times\mathbf{B},\]
 y la fuerza sobre la part\'{\i}cula resulta ser\begin{equation}
\mathbf{F}=\nabla(\boldsymbol{\mu}\cdot\mathbf{B}).\label{eq0}\end{equation}
 En los experimentos (véase más abajo) se suele disponer un campo inhomogéneo
$\mathbf{B}=(B_{x},0,B_{z})$, en la región por la que va a pasar
el haz de part\'{\i}culas, se ajustan los parámetros del campo de
manera que se tenga $B_{z}\gg B_{x}$ y as\'{\i} $\mathbf{B}\simeq B_{z}\mathbf{\hat{z}}$.
En estas condiciones, de la igualdad $\boldsymbol{\tau}=\frac{d\mathbf{S}}{dt}$
resulta \begin{align*}
\frac{dS_{x}}{dt} & =\gamma S_{y}B_{z}\\
\frac{dS_{y}}{dt} & =-\gamma S_{x}B_{z}\\
\frac{dS_{z}}{dt} & =0.\end{align*}
 Es decir, si escribimos $\mathbf{S}=\mathbf{S}_{||}+\mathbf{S}_{\bot}$,
donde $\mathbf{S}_{||}$ es la componente de $\mathbf{S}$ paralela
al eje $\mathbf{\hat{z}}$ y $\mathbf{S}_{\bot}$ su componente perpendicular,
tenemos que la primera es constante, mientras que la segunda gira
alrededor del eje $\mathbf{\hat{z}}$ con una velocidad angular $\omega=\gamma B_{z}$.

Por otra parte, la componente de la fuerza que actúa sobre las part\'{\i}culas
del haz en la dirección del eje $\mathbf{\hat{z}}$ es\[
F_{z}=\frac{\partial}{\partial z}\left(\mu_{z}B_{z}+\mu_{x}B_{x}\right)\simeq\frac{\partial}{\partial z}\left(\mu_{z}B_{z}\right)=\mu_{z}\frac{\partial B_{z}}{\partial z},\]
 por lo que se ve claramente que $F_{z}$ produce una desviación de
las part\'{\i}culas en la dirección $\mathbf{\hat{z}}$ proporcional
a la proyección $\mu_{z}$. Conociendo $\frac{\partial B_{z}}{\partial z}$
y esta desviación, es posible conocer $\mu_{z}$ y, obviamente, no
hay motivo alguno para no esperar medir todos los valores posibles
de esta componente, que irán desde $-\left\vert \mathbf{\mu}\right\vert $
a $\left\vert \mathbf{\mu}\right\vert $.

\begin{figure}[ht]
\centering%
\includegraphics[width=10cm]{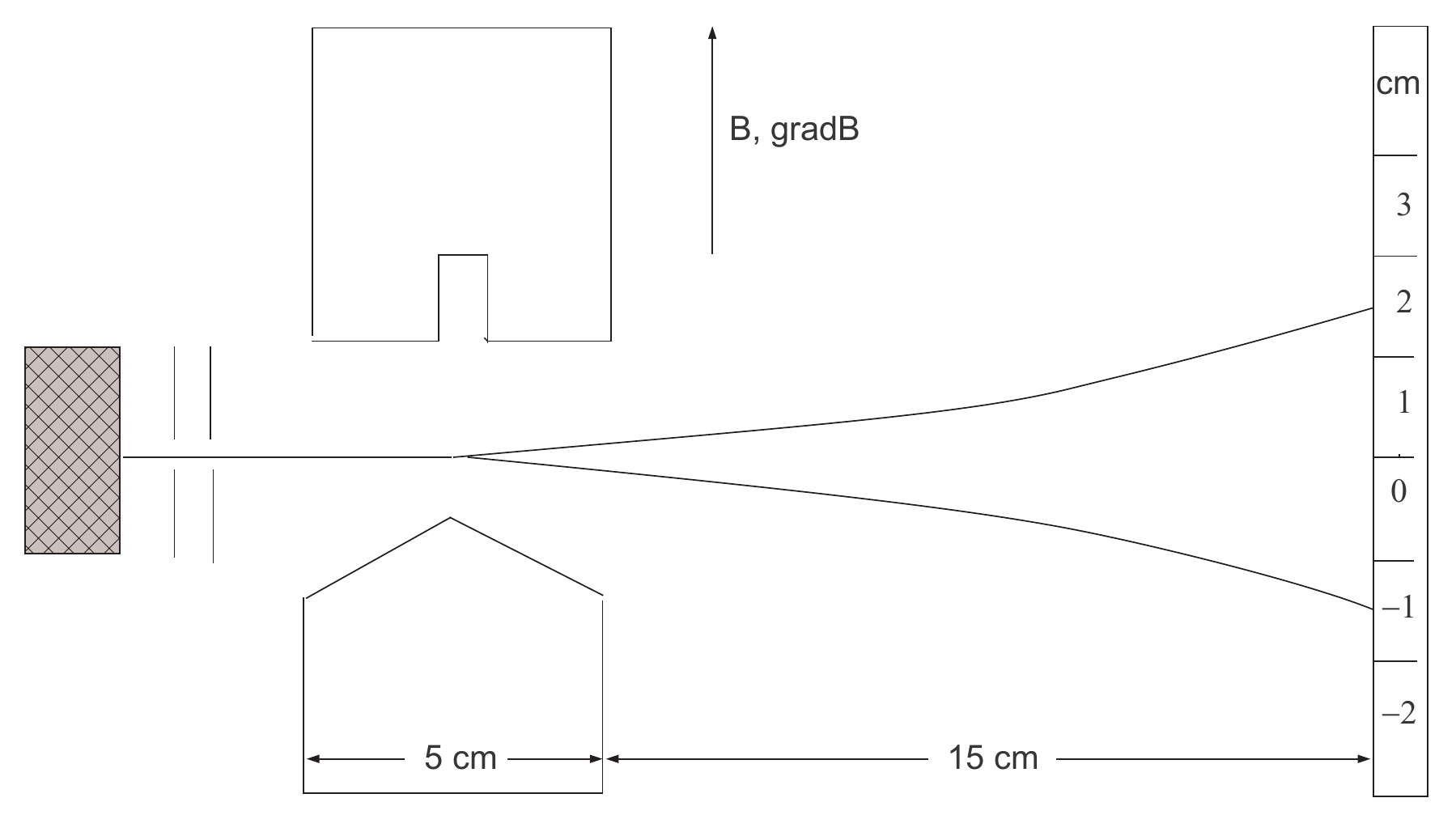}
\caption{Experimento de Stern-Gerlach.}
\label{fig1} 
\end{figure}

Estas predicciones estaban siendo verificadas por O. Stern y W. Gerlach
utilizando un haz de átomos de plata, cuando se encontraron con la
sorpresa de que el comportamiento observado no era el esperado (ver
\cite{Ger-Ste 24}): se observaba una desviación, pero no hacia arriba
o hacia abajo, sino hacia arriba y abajo a la vez (la Figura \ref{fig1}
da una representación esquemática del experimento).

La magnitud de la desviación era la misma en ambos rayos y esto era
totalmente incompatible con el modelo clásico. En 1925 S. Goudsmith
y G. Uhlenbeck (ver \cite{Gou-Uhl 25}) propusieron una solución al
enigma planteado por este experimento mediante la introducción de
un nuevo ``número cuántico''%
\footnote{Históricamente, Goudsmith y Uhlenbeck no introdujeron la hipótesis
del espín para explicar los resultados de Stern-Gerlach, sino para
corregir unas irregularidades observadas en el estudio de ciertas
series espectroscópicas, pero pronto se pudo ver, a la luz de la nueva
propuesta, que la explicación que hasta el momento se daba del experimento
de Stern-Gerlach (basadas en la llamada {}``cuantización espacial\textquotedblright
) era errónea, y que el espín serv\'{\i}a también para dar una nueva
teor\'{\i}a más acertada.%
}, el \emph{espín}, capaz de interaccionar con un campo magnético. De
hecho, formalmente las propiedades del espín son las mismas que las
de un momento angular ordinario, por lo que a veces se le denomina
``momento angular interno''\ (sin embargo, debe resaltarse el
hecho de que el espín no está asociado a ningún movimiento espacial,
ni de la part\'{\i}cula ni de su distribución de carga asociada, es
una magnitud {}``intr\'{\i}nseca''\ a cada part\'{\i}cula, de ah\'{\i}
el calificativo de momento {}``interno'' ). Al igual que al momento
angular orbital asociado al movimiento de rotación del electrón alrededor
del núcleo se le asocia un momento magnético, al nuevo momento angular
de espín se le asocia un momento magnético $\mathbf{m}$, con las mismas
propiedades que $\boldsymbol{\mu}$, en particular, su forma de acoplarse
a un campo magnético.

El caso es que experimentalmente se determina que el espín del electrón 
vale $1/2$, y el operador (observable)
correspondiente proyectado sobre un eje espacial, digamos el del campo
magnético entre los imanes $\mathbf{B}$, admite sólo dos valores
propios: $1/2$ y $-1/2$ (véase la sección \ref{spin1/2}). Esto,
junto con el hecho de que el momento angular neto de los átomos de
plata se debe únicamente al momento de espín del electrón exterior
de su capa de valencia, explica por qué se observan dos rayos desviados
en sentidos opuestos%
\footnote{Un átomo de plata tiene $47$ electrones, de los cuales $46$ se encuentran
en un estado que tiene momento angular orbital y de espín nulos, mientras
que el electrón restante se encuentra en un estado $s^{1}$, que también
tiene momento angular orbital nulo; por tanto, el espín total del átomo
es el momento de espín de este electrón%
}.

\section{Espacio de estados y observables\label{estados}}

La descripción cuántica habitual del estado de un sistema f\'{\i}sico
se lleva a cabo a través de su función de onda asociada, un elemento
$\psi\in L^{2}(\mathbb{R}^{3})$ del espacio de las funciones con
valores complejos y de cuadrado integrable (en sentido Lebesgue) de
$\mathbb{R}^{3}$. Este espacio resulta ser de Hilbert, y su producto
escalar está dado por \begin{equation}
\left\langle \psi,\varphi\right\rangle =\int\nolimits _{\mathbb{R}^{3}}\psi^{\ast}\varphi,\label{eq0_0}\end{equation}
 donde $\psi^{\ast}$ denota la conjugada compleja de $\psi$.

Realmente, como estamos interesados en estudiar evoluciones temporales,
los objetos básicos son curvas $f:\mathbb{R}\rightarrow L^{2}(\mathbb{R}^{3})$
tales que $f(t)=\psi_{t}(\mathbf{x})$, y es común escribir simplemente
$\psi_{t}(\mathbf{x})\equiv\psi(t,\mathbf{x})$ diciendo que {}``la
función de onda depende del tiempo'' .

\textit{\emph{Una observación}}: no hay que confundir $\psi_{t}(\mathbf{x})\equiv\psi(t,\mathbf{x})$
con una trayectoria de la partícula. La interpretación estándar de
$\psi_{t}(\mathbf{x})$ sólo afirma que en cada instante $t$ hay
una cierta probabilidad de hallar la partícula en un entorno de $\mathbf{x}$,
no que en su movimiento la partícula describa una trayectoria definida
(tal concepto está fuera de la Mecánica Cuántica debido a las relaciones
de incertidumbre de Heisenberg, véase el Apéndice A).

De acuerdo con la interpretación probabilista de Max Born, el significado
f\'{\i}sico de la función de onda es que la probabilidad de encontrar
el sistema en una posición e instante determinados viene descrita
por la \emph{función de densidad} \begin{equation}
\rho(t,\mathbf{x})=\psi^{\ast}(t,\mathbf{x})\cdot\psi(t,\mathbf{x}),\label{eq0_01}\end{equation}
 supuesta la normalización \begin{equation}
\left\Vert \psi_{t}(\mathbf{x})\right\Vert _{L^{2}(\mathbb{R}^{3})}^{2}=\int\nolimits _{\mathbb{R}^{3}}\psi_{t}^{\ast}\psi_{t}=1.\label{eq0_02}\end{equation}

La interpretación de Copenhague de la Mecánica Cuántica, basada en
este postulado de Born, afirma además que el producto (\ref{eq0_0})
proporciona la \emph{amplitud de probabilidad de transición} desde
un estado del sistema descrito por $\varphi$ a otro descrito por
$\psi$.

En la notación de Dirac, un elemento de $L^{2}(\mathbb{R}^{3})$ tal
como $\varphi$ (ó $\varphi(t,\mathbf{x})$ si trabajamos con curvas)
se escribe $\left\vert \varphi\right\rangle $ (ó $\left\vert \varphi(t,\mathbf{x})\right\rangle $)
y se denomina {}``ket'' . De acuerdo con el Teorema de Representación
de Riesz, si $\psi$ es otro elemento de $L^{2}(\mathbb{R}^{3})$
o un conjunto de ellos dados por una curva $\psi(t,\mathbf{x})$,
a cada uno le corresponde un elemento del dual topológico $\left(L^{2}(\mathbb{R}^{3})\right)^{\prime}$,
es decir, un funcional continuo que actúa mediante el producto escalar
y que se representa por $\left\langle \psi(t,\mathbf{x})\right\vert $,
el llamado {}``bra'': para cada $t\in\mathbb{R}$ se tiene\[
\begin{array}{llll}
\left\langle \psi(t,\mathbf{x})\right\vert  & :L^{2}(\mathbb{R}^{3}) & \rightarrow & \mathbb{C}\\
 & \left\vert \varphi(t,\mathbf{x})\right\rangle  & \mapsto & \left\langle \psi(t,\mathbf{x}),\varphi(t,\mathbf{x})\right\rangle .\end{array}\]
 As\'{\i}, para pasar de un estado $\left\vert \varphi(t,\mathbf{x})\right\rangle $
a otro $\left\vert \psi(t,\mathbf{x})\right\rangle $ no hay más que
formar el {}``bracket'' de ambos.

Resulta evidente que la función de densidad (\ref{eq0_01}), que codifica
la información básica acerca de la part\'{\i}cula, no se ve alterada
si en lugar de $\psi(t,\mathbf{x})$ tomamos su modificación por un
factor de fase global, esto es, una función de onda de la forma $e^{i\alpha}\psi(t,\mathbf{x})$,
donde $\alpha\in\mathbb{C}$. Y, de hecho, ya hemos utilizado esta
caracter\'{\i}stica al imponer la normalización (\ref{eq0_02}); en
este sentido, a veces se dice que los estados de un sistema f\'{\i}sico
están descritos por rayos en el espacio $L^{2}(\mathbb{R}^{3})$:
se toma como espacio de trabajo el cociente $L^{2}(\mathbb{R}^{3})\diagup\mathcal{R}$,
donde la relación de equivalencia es: $\psi\mathcal{R}\varphi$ si
y sólo si existe un $\alpha\in\mathbb{C}$ tal que $\psi=e^{i\alpha}\varphi$
(es decir, se trabaja con el espacio proyectivizado $P(L^{2}(\mathbb{R}^{3}))$).
Entonces, el normalizar la función de onda equivale a trabajar con
el representante de la clase con norma unitaria.

Sobre estos estados $\left\vert \psi(t,\mathbf{x})\right\rangle $
actúan los \emph{observables}: llamamos observable a un operador
$A\in\mathrm{End}_{\mathbb{C}}L^{2}(\mathbb{R}^{3})$ tal que es autoadjunto respecto
del producto escalar en $L^{2}(\mathbb{R}^{3})$. Es sabido que, en
tal caso, $A$ tiene todos sus valores propios reales y esos valores
propios (según postula la Mecánica Cuántica) son los posibles valores
de las mediciones de (la magnitud f\'{\i}sica que representa) $A$
sobre el sistema.

La forma habitual de construir estos observables cuánticos procede
por analog\'{\i}a con sus contrapartidas clásicas. Sin embargo, no
existe un método general para realizar este proceso, conocido como
\emph{cuantización}, por lo que definirlos
correctamente es casi un arte. De hecho, hay propiedades observables
cuánticas que no tienen un análogo clásico directo (como es el caso
del espín en el experimento de Stern-Gerlach) y, en definitiva, el
problema de la cuantización es fundamental y sigue abierto.

En cualquier caso, lo que nos interesa resaltar ahora es que si sólo
consideramos operadores actuando sobre funciones (o curvas) en $L^{2}(\mathbb{R}^{3})$,
nos estamos restringiendo a propiedades del sistema que dependen de
su extensión espacio-temporal, pero no es posible describir entonces
propiedades ``intr\'{\i}nsecas'' como el espín de Goudsmith y Uhlenbeck.
Es preciso ampliar el espacio de estados de $L^{2}(\mathbb{R}^{3})$
a otro espacio de Hilbert en el que tengan cabida los objetos necesarios
para describir propiedades como el espín, que no dependen del movimiento
del sistema en el espacio f\'{\i}sico tridimensional%
\footnote{Que esto es as\'{\i} puede verse considerando que los resultados del
experimento de Stern-Gerlach no dependen de la orientación espacial
del dispositivo ni del estado de movimiento de éste.%
}.

En los siguientes párrafos veremos cómo la descripción del espín $1/2$
(es decir, una magnitud f\'{\i}sica representada por un operador autoadjunto
que sólo admite dos valores propios, uno positivo y otro negativo)
puede hacerse muy fácilmente en un espacio bien sencillo: $\mathbb{C}^{2}=\mathbb{C}\times\mathbb{C}$.
Si a este espacio lo llamamos ``espacio de espín'', por contraposición
al ``espacio de grados de libertad espacio-temporales $L^{2}(\mathbb{R}^{3})$''
, vemos que el espacio total de estados para una part\'{\i}cula de
espín $1/2$ es\[
\mathcal{H}=L^{2}(\mathbb{R}^{3})\otimes\mathbb{C}^{2},\]
 y los observables f\'{\i}sicos serán los endomorfismos autoadjuntos
de $\mathcal{H}$. La evolución del sistema nuevamente estará descrita
por curvas en $\mathcal{H}$: si $f:\mathbb{R}\rightarrow\mathcal{H}$
es una de ellas, se tendrá\begin{equation}
f(t)=\psi\otimes\left(\begin{array}{l}
\alpha\\
\beta\end{array}\right)\equiv\left\vert \psi(t,\mathbf{x})\right\rangle \otimes\left(\begin{array}{l}
\alpha(t)\\
\beta(t)\end{array}\right),\label{eq0_002}\end{equation}
(con la particularidad de que puede darse el caso en que $\alpha(t)$ y $\beta(t)$
sean funciones constantes).

Por supuesto esta no es la única posibilidad, hay otras y para un
espín diferente se necesita otro espacio, pero para la discusión del
modelo supersimétrico sencillo que presentaremos esta descripción
es suficiente.

\section{Descripción de part\'{\i}culas con espín $1/2$\label{spin1/2}}

¿Cómo explicar el resultado del experimento de Stern-Gerlach?. De
acuerdo con la hipótesis de Goudsmith y Uhlenbeck, el fenómeno observado
se debe a que el electrón tiene espín $1/2$ y a que esta nueva magnitud
tiene dos posibles {}``proyecciones'', $\pm\frac{1}{2}$. Un modelo
para esta situación es el siguiente: consideremos el espacio\[
\mathbb{C}^{2}=\left\{ \eta=\left(\begin{array}{l}
\alpha\\
\beta\end{array}\right):\alpha,\beta\in\mathbb{C}\right\} ,\]
 dotado del producto escalar complejo (herm\'{\i}tico)\[
\left\langle \eta,\zeta\right\rangle =\left\langle \left(\begin{array}{l}
\alpha\\
\beta\end{array}\right),\left(\begin{array}{l}
\gamma\\
\delta\end{array}\right)\right\rangle =\alpha^{\ast}\gamma+\beta^{\ast}\delta=\eta^{\dag}\zeta,\]
 con $\eta^{\dag}=\left(\eta^{\ast}\right)^{\top}$ la conjugación
herm\'{\i}tica (la traspuesta de la matriz conjugada compleja). Diremos
que $\eta\in\mathbb{C}^{2}$ es un \emph{estado de espín} si $\left\Vert \eta\right\Vert =\sqrt{\left\langle \eta,\eta\right\rangle }=1$.
Identificaremos dos estados $\eta,\zeta$ si existe un $z\in\mathbb{C}$
con $\left\vert z\right\vert =1$ tal que \[
\eta=z\cdot\zeta=\left(\begin{array}{l}
z\gamma\\
z\delta\end{array}\right).\]
 Es decir, si llamamos $\sim$ a esta relación de equivalencia y $\mathcal{U}(\mathbb{C}^{2})=\{\eta\in\mathbb{C}^{2}:\left\Vert \eta\right\Vert =1\}$,
trabajaremos con el espacio cociente\[
\mathcal{S}_{2}=\mathcal{U}(\mathbb{C}^{2})\diagup\sim\]
 cuyos elementos se denominan \emph{espinores bidimensionales} o \emph{biespinores}.
Sin embargo, por comodidad, denotaremos las clases de equivalencia
$[\eta]_{\sim}\in\mathcal{S}_{2}$ por sus representantes, es decir,
escribiremos $[\eta]_{\sim}=\eta$.

¿Cuál es la estructura del espacio $\mathcal{S}_{2}$?. Observemos
que\[
\mathcal{S}_{2}=\{\eta\in\mathbb{C}^{2}\simeq\mathbb{R}^{4}:\left\Vert \eta\right\Vert =1\}\simeq S^{3}\]
 (la esfera unitaria en $\mathbb{R}^{4}$), y que la relación de equivalencia
$\sim$, que determina los estados f\'{\i}sicamente equivalentes,
nos dice que se identifican todos los puntos que bajo la correspondencia
anterior van a parar a $S^{1}$ (la esfera unitaria en $\mathbb{R}^2$). Más concretamente, la órbita de un
elemento cualquiera en el cociente $\mathcal{S}_{2}$ es difeomorfa
a $S^{1}$, pues\[
[\eta]_{\sim}=\{\zeta\in\mathbb{C}^{2}:\exists\left\vert z\right\vert =1,\zeta=z\cdot\eta\}\equiv S^{1}\cdot\eta\text{.}\]
 Por tanto, el espacio cociente (o espacio de órbitas) es\[
\mathcal{S}_{2}=S^{3}\diagup S^{1}\simeq S^{2}\text{,}\]
(donde $S^{2}$ es la esfera unitaria en $\mathbb{R}^3$)
 que resulta, por tanto, ser bidimensional. Como curiosidad, mencionaremos
que esta construcción es equivalente a lo que en Matemáticas se conoce
como fibración de Hopf de $S^{3}$.

Llamaremos \emph{observables del espacio de espín} a los $\mathbb{C-}$endomorfismos
autoadjuntos de $\left(\mathbb{C}^{2},\left\langle .,.\right\rangle \right)$,
esto es, a las aplicaciones $\mathbb{C\mathnormal{-}}$lineales autoadjuntas
$A:\mathbb{C}^{2}\rightarrow\mathbb{C}^{2}$. Del Álgebra Lineal,
es sabido que una tal aplicación puede verse como una matriz $A\in\mathrm{Mat}_{2\times2}(\mathbb{C})$
con coeficientes complejos:
\[
A=\left(\begin{array}{ll}
\alpha & \beta\\
\gamma & \delta\end{array}\right)
\]
 tal que $A^{\dag}=A$ (se dice a veces que $A$ es una matriz herm\'{\i}tica,
y se representa por $A\in\mathrm{Her}_{2\times2}(\mathbb{C})$). En
particular, esto implica que $A$ posee dos valores propios reales
$\lambda_{1},\lambda_{2}\in\mathbb{R}$ (se escribe entonces $\lambda_{1},\lambda_{2}\in\mathrm{Spec}_{\mathbb{R}}(A)$).
Además, dado uno de estos observables existen $\eta_{1},\eta_{2}\in\mathbb{C}^{2}$
vectores propios tales que
\begin{eqnarray*}
A\eta_{1} & = & \lambda_{1}\cdot\eta_{1}\\
A\eta_{2} & = & \lambda_{2}\cdot\eta_{2},
\end{eqnarray*}
 y $\{\eta_{1},\eta_{2}\}$ es una \emph{base ortonormal} compleja
de $\mathbb{C}^{2}$.

Como su propio nombre indica, los observables están asociados a medidas
sobre el sistema f\'{\i}sico que representan los estados. Veamos cómo
son los resultados de estas medidas: supongamos que un sistema se
encuentra en el estado descrito por el vector $\zeta\in\mathbb{C}^{2}$
y que se realiza una medida correspondiente al observable $A$, con
$\mathrm{Spec}_{\mathbb{R}}(A)=\{\lambda_{1},\lambda_{2}\}$ y vectores
propios $\{\eta_{1},\eta_{2}\}$. Entonces, se puede escribir
\[
\zeta=\alpha_{1}\cdot\eta_{1}+\alpha_{2}\cdot\eta_{2},
\]
 donde (por ser $\{\eta_{1},\eta_{2}\}$ ortonormal)
 \[
\left\vert \alpha_{1}\right\vert ^{2}+\left\vert \alpha_{2}\right\vert ^{2}=1.
\]
 Pues bien, la hipótesis fundamental de la interpretación estándar
de la Mecánica Cuántica dice que la medición del observable representado
por $A$ en el estado $\zeta$ da como posibles resultados $\lambda_{1},\lambda_{2}$
con probabilidades respectivas $\left\vert \alpha_{1}\right\vert ^{2},\left\vert \alpha_{2}\right\vert ^{2}$
y, además, tras la medición, si se ha obtenido el resultado $\lambda_{j}$
$(j=1,2)$ el sistema pasa a estar descrito por el estado $\eta_{j}$
$(j=1,2)$; a este fenómeno se le conoce como \emph{colapso de la
función de onda}. Dentro de este esquema, $\alpha_{j}=\left\langle \eta_{j},\zeta\right\rangle $
es la amplitud de probabilidad para la transición del estado $\zeta$
al $\eta_{j}$, y $\left\vert \alpha_{j}\right\vert ^{2}$ es la probabilidad
de que esto se produzca.

A $\left\langle \eta\left\vert A\right\vert \eta\right\rangle =\left\langle \eta,A\eta\right\rangle \in\mathbb{R}$
se le llama \emph{valor esperado} del observable $A$ en el estado
$\eta$. Observemos que
\[
\left\langle \eta\left\vert A\right\vert \eta\right\rangle =\left\vert \alpha_{1}\right\vert ^{2}\lambda_{1}+\left\vert \alpha_{2}\right\vert ^{2}\lambda_{2},
\]
 de modo que el valor esperado de $A$ oscila entre $\lambda_{1}$
y $\lambda_{2}$. También, fijémonos en que los valores esperados
de un observable son independientes del representante elegido para
describir el estado del sistema, ya que dos de ellos se relacionan
por un $z\in\mathbb{C}$ tal que $\left\vert z\right\vert^2 =z^{\ast}z=1$:
si $\eta$ y $\zeta=z\cdot\eta$ son dos representantes de un mismo
estado f\'{\i}sico
\[
\left\langle \zeta\left\vert A\right\vert \zeta\right\rangle =\left\langle z\cdot\eta,A(z\cdot\eta)\right\rangle =z^{\ast}z\left\langle \eta\left\vert A\right\vert \eta\right\rangle =\left\langle \eta\left\vert A\right\vert \eta\right\rangle \text{.}
\]

Al igual que hicimos con los estados, ocupémonos ahora de la estructura
del espacio de los observables. Si $A\in\mathrm{Mat}_{2\times2}(\mathbb{C})$
es uno de ellos, debe cumplir la condición $A^{\dag}=A$, de modo
que si su representación matricial es
\[
A=\left(\begin{array}{ll}
\alpha & \beta\\
\gamma & \delta\end{array}\right),
\]
 como
 \[
A^{\dag}=\left(\begin{array}{ll}
\alpha^{\ast} & \gamma^{\ast}\\
\beta^{\ast} & \delta^{\ast}\end{array}\right),
\]
 debe verificarse
 \[
\begin{array}{l}
\alpha=\alpha^{\ast}=r\in\mathbb{R}\\
\delta=\delta^{\ast}=s\in\mathbb{R}\\
\beta=\gamma^{\ast}=u+iv\in\mathbb{C,}\end{array}
\]
 o sea,
 \[
A=\left(\begin{array}{lr}
r & u+iv\\
u-iv & s\end{array}\right).
\]
 Por motivos que serán evidentes enseguida, nos interesará reparametrizar
$A$, poniendo en lugar de $r,s,u,v\in\mathbb{R}$ unos nuevos coeficientes
$a_{0},a_{1},a_{2},a_{3}$ relacionados con los anteriores por
\[
\begin{array}{l}
a_{0}+a_{3}=r\\
a_{0}-a_{3}=s\\
a_{1}=u\\
a_{2}=v,\end{array}
\]
 as\'{\i} que nuestro observable más general $A$ pasa a ser
 \begin{equation}
A=\left(\begin{array}{lr}
a_{0}+a_{3} & a_{1}+ia_{2}\\
a_{1}-ia_{2} & a_{0}-a_{3}\end{array}\right).\label{eq0_03}
\end{equation}

Resulta entonces que el espacio de todos los observables puede verse
como\[
\mathrm{Obs}=\{A\in\mathrm{End}\left(\mathbb{C}^{2},\left\langle .,.\right\rangle \right):A=A^{\dag}\}\simeq\mathbb{R}^{4}\text{,}\]
 que, a su vez, tiene estructura de espacio vectorial \emph{real}
y de (\ref{eq0_03}) se puede hallar fácilmente una base expl\'{\i}cita.
De hecho, si se consideran las llamadas \emph{matrices de Pauli}
\[
\sigma_{1}=\left(\begin{array}{lr}
0 & 1\\
1 & 0\end{array}\right),\text{ }\sigma_{2}=\left(\begin{array}{lr}
0 & -i\\
i & 0\end{array}\right),\text{ }\sigma_{3}=\left(\begin{array}{lr}
1 & 0\\
0 & -1\end{array}\right),
\]
 y se escribe $\sigma_{0}=\left(\begin{array}{lr}
1 & 0\\
0 & 1\end{array}\right)=\mathbf{I}$ para la matriz identidad, un observable $A$ se expresa como 
\[
A=a_{0}\sigma_{0}+a_{1}\sigma_{1}+a_{2}\sigma_{2}+a_{3}\sigma_{3}=a_{0}\mathbf{I}+\mathbf{a\cdot\sigma,}
\]
 donde $\mathbf{\sigma}=(\sigma_{1},\sigma_{2},\sigma_{3}),\mathbf{a}=(a_{1},a_{2},a_{3})$
(esto es sólo una notación cómoda).

Ahora, es muy fácil comprobar%
\footnote{Ejercicio para el lector} que si se expresa $A$ a partir de los parámetros $a_{0},\mathbf{a}$,
sus dos posibles valores propios son 
\begin{equation}
\lambda_{\pm}=a_{0}\pm\sqrt{a_{1}^{2}+a_{2}^{2}+a_{3}^{2}}=a_{0}\pm\left\Vert \mathbf{a}\right\Vert .
\label{eq0_04}
\end{equation}
 Si $\left\Vert \mathbf{a}\right\Vert =0$, los valores propios son
degenerados e iguales a $a_{0}$.

Por definición, el operador \emph{observable de espín} $1/2$ (o \emph{momento
de espín} $1/2$) es el vector de operadores autoadjuntos $S$ dado por
$\mathbf{S}=\frac{1}{2}(\sigma_{1},\sigma_{2},\sigma_{3})$.
Generalizando la descripción clásica de la sección
\ref{SG}, una part\'{\i}cula con momento de espín $\frac{1}{2}$ se
comporta como si tuviera un momento magnético asociado a ese espín,
de la misma forma que tiene asociado un mometo magnético a su momento
angular orbital: se introduce pues un \emph{momento magnético asociado
al espín} $\mathbf{m}$, dado por\[
\mathbf{m}=\mu_{B}\mathbf{S},\]
es decir, el nuevo operador es 
$\mathbf{m}=\frac{\mu_{B}}{2}(\sigma_{1},\sigma_{2},\sigma_{3})$,
donde $\mu_{B}$ es una constante que se conoce como \emph{magnetón
de Bohr}. Lo que interesa, por lo que respecta a la fuerza que actúa
sobre la part\'{\i}cula, es conocer los valores propios de la proyección
de $\mathbf{m}$ sobre una dirección dada en el espacio (cfr. (\ref{eq0})):
si $\mathbf{u}=(u_{1},u_{2},u_{3})\in\mathbb{R}^{3}$ es una tal dirección,
resulta que la proyección del operador $\mathbf{m}$ es 
\[
\mathbf{m\cdot u}=\frac{\mu_{B}}{2}(u_{1}\cdot\sigma_{1}+\sigma_{2}\cdot u_{2}+\sigma_{3}\cdot u_{3})=\frac{\mu_{B}}{2}\left(\begin{array}{lr}
u_{3} & u_{1}-iu_{2}\\
u_{1}+iu_{2} & -u_{3}
\end{array}\right),
\]
 y de acuerdo con (\ref{eq0_04}), los valores propios de esta proyección
son\[
\lambda_{\pm}=\pm\frac{\mu_{B}}{2}\left\Vert \mathbf{u}\right\Vert ,\]
 as\'{\i} que si $\mathbf{u}$ es unitario obtenemos $\lambda_{\pm}=\pm\frac{\mu_{B}}{2}$.
Es habitual tomar $\mathbf{u}=\mathbf{\hat{z}}$, en cuyo caso se
escribe\[
m_{z}=\pm\frac{\mu_{B}}{2}\]
 para los valores propios de la proyección del momento magnético de
espín.

Por tanto, este formalismo da cuenta de los resultados del experimento
de Stern-Gerlach: en un haz de átomos de plata, los electrones ``exteriores''
cuyo espín es $1/2$ se hallarán en uno de los dos estados posibles
para la proyección del espín, el dado por el valor propio $\frac{\mu_{B}}{2}$
o el dado por $-\frac{\mu_{B}}{2}$, y las fuerzas que actúan en cada
caso serán colineales pero de sentidos opuestos, es decir, los átomos
se desviarán en uno de los dos sentidos (arriba o abajo) dependiendo
de si el estado del electrón ``exterior'' se corresponde con el
vector propio correspondiente a $\frac{\mu_{B}}{2}$ o $-\frac{\mu_{B}}{2}$.

Por otra parte, tenemos que los estados del espacio de espín $\mathbb{C}^{2}$
se pueden expresar en la base ortonormal formada por los vectores
propios de $\mathbf{m\cdot u},$ llamémoslos $\eta_{\pm}$, y que
están definidos por
\[
\left(\mathbf{m\cdot u}\right)\eta_{\pm}=\lambda_{\pm}\eta_{\pm}.
\]
 As\'{\i}, para todo $\zeta\in\mathbb{C}^{2}$ se puede poner $\zeta=\alpha_{1}\eta_{+}+\alpha_{2}\eta_{-}$,
con $\left\vert \alpha_{1}\right\vert ^{2}+\left\vert \alpha_{2}\right\vert ^{2}=1$.
Como ya hemos mencionado, es común tomar $\mathbf{u}=\mathbf{\hat{z}}$
y escribir $\eta_{+}=\eta_{\uparrow},\eta_{-}=\eta_{\downarrow}$(con
sus correspondientes $\alpha_{1}=\alpha_{\uparrow},\alpha_{-}=\alpha_{\downarrow}$),
de modo que un estado de espín será lo que se conoce como un ``biespinor'':
\[
\zeta=\left(\begin{array}{l}
\alpha_{\uparrow}\\
\alpha_{\downarrow}
\end{array}\right)=\alpha_{\uparrow}\eta_{\uparrow}+\alpha_{\downarrow}\eta_{\downarrow},
\]
 y las funciones de onda completas (\ref{eq0_002}) se escriben entonces
en la forma 
\begin{equation}
\begin{array}{r}
f(t)=\left\vert \psi(t,\mathbf{x})\right\rangle \otimes\left(\alpha_{+}\left\vert +\right\rangle +\alpha_{-}\left\vert -\right\rangle \right)\\
\\=\alpha_{+}\left\vert t,\mathbf{x},+\right\rangle +\alpha_{-}\left\vert t,\mathbf{x},-\right\rangle 
\end{array}\label{eq0_04a}
\end{equation}
donde, obviamente, $\left\vert t,\mathbf{x},\pm\right\rangle =\left\vert \psi(t,\mathbf{x})\right\rangle \otimes\left\vert \pm\right\rangle $.

\section{Las ecuaciones de Schrödinger y Dirac-Heisenberg\label{S5}}

Al igual que la dinámica en Mecánica Clásica viene determinada por
la ecuación de Newton para una trayectoria
en el espacio euclideo tridimensional $\mathbf{x}=\mathbf{x}(t)$,
que para fuerzas que derivan de un potencial se escribe
en función del operador gradiente $\nabla$ como
\[
\frac{d^{2}\mathbf{x}}{dt^{2}}=-\nabla V\circ\mathbf{x},\]
en Mecánica Cuántica toda part\'{\i}cula interaccionando con un potencial
$V(\mathbf{x})$ (supongamos por simplicidad potenciales dependientes sólo
de la posición) viene descrita por su función de onda $\Psi(\mathbf{x},t)$
\footnote{Para explicar la aparición de $t$ recordemos que, realmente, cuando
decimos ``función de onda $\Psi(\mathbf{x},t)$'' nos estamos refiriendo a una ``curva en el
espacio de estados $\Psi :\mathbb{R} \rightarrow L^{2}(\mathbb{R}^3) \otimes \mathbb{C}^{2}$''.
La ecuación de Schrödinger, como ecuación dinámica, es una ecuación para estas curvas
pero el uso ha consagrado la identificación de la curva $\Psi$ con su imagen.
},
que satisface la llamada \emph{ecuación de Schrödinger} 
\begin{equation}
i\hbar\frac{\partial\Psi}{\partial t}=\left(-\frac{\hbar^{2}}{2m}
\left( 
\frac{\partial^{2}\Psi}{\partial x^{2}}
+ \frac{\partial^{2}\Psi}{\partial y^{2}}
+ \frac{\partial^{2}\Psi}{\partial z^{2}}
\right)
+V(\mathbf{x})
\right)\Psi(\mathbf{x},t)\equiv H\Psi(\mathbf{x},t),
\label{eq0_04b}
\end{equation}
 aqu\'{\i}, $H=-\frac{\hbar^{2}}{2m}
 \left(\frac{\partial^{2}\cdot}{\partial x^{2}}+
 \frac{\partial^{2}\cdot}{\partial y^{2}}+
 \frac{\partial^{2}\cdot}{\partial z^{2}}\right)+
 V(\mathbf{x})\cdot$
es el observable denominado \emph{Hamiltoniano cuántico del sistema}.
Aplicando el conocido método de separación de variables (es decir,
suponiendo que $\Psi(\mathbf{x},t)=\Phi(\mathbf{x})F(t)$), esta ecuación puede reducirse
a otra más sencilla, denominada \emph{ecuación de Schrödinger independiente
del tiempo}:
\begin{equation}
H\Phi(\mathbf{x})=E\Phi(\mathbf{x}),\label{eq0_05}
\end{equation}
 donde $E$ es una constante real%
\footnote{\label{notaapie1}Observemos que $E$ resulta ser un valor propio
del observable $H$ y éste, por hipótesis, es un operador autoadjunto.%
} que, por analog\'{\i}a con el caso clásico, se identifica con la
energ\'{\i}a del sistema. Esta ecuación puede generalizarse al caso
de que existan ``variables internas'' como el espín; de hecho,
recordando que el espacio de Hilbert que describe los estados del
sistema (por ejemplo para una part\'{\i}cula de espín $1/2$) tiene la 
estructura $\mathcal{H}=L^{2}(\mathbb{R}^3)\otimes\mathbb{C}^{2}$,
lo anterior deber\'{\i}a escribirse: 
\[
\left(i\hbar\frac{\partial}{\partial t}\otimes\frac{\partial}{\partial t}\right)\Psi(x,t)=
\left(\left(-\frac{\hbar^{2}}{2m}\left(
\frac{\partial^{2}}{\partial x^{2}}+ \frac{\partial^{2}}{\partial y^{2}}+ 
\frac{\partial^{2}}{\partial z^{2}} 
\right)+
V_{L^{2}(\mathbb{R})}\right)\otimes V_{\mathbb{C}^{2}}\right)\Psi(x,t),
\]
 esto es, el Hamiltoniano ser\'{\i}a 
 \[
H=\left(-\frac{\hbar^{2}}{2m}
\left(
\frac{\partial^{2}}{\partial x^{2}}+\frac{\partial^{2}}{\partial y^{2}}+ 
\frac{\partial^{2}}{\partial z^{2}}
\right)+V_{L^{2}(\mathbb{R})}\right)\otimes V_{\mathbb{C}^{2}}\text{.}
\]
 En concreto, para una part\'{\i}cula de espín $1/2$ en el seno de
un campo magnético $\mathbf{B}$, ya hemos visto que es
\[
V_{\mathbb{C}^{2}}=\mu_{B}\mathbf{S}\cdot\mathbf{B},
\]
 y as\'{\i}, cuando sólo interesa estudiar la evolución del factor
de la función de onda completa pertenenciente a $\mathbb{C}^{2}$,
se dice que el Hamiltoniano de una part\'{\i}cula de espín $1/2$ en
un campo magnético $\mathbf{B}$ es
\[
H=\mu_{B}\mathbf{S}\cdot\mathbf{B}.
\]

Por tanto, vemos que es el Hamiltoniano el que determina toda la dinámica
del sistema, que básicamente se reduce a un problema de valores propios
para este operador.

Heisenberg y, de manera independiente, Dirac proporcionaron una ecuación
básica para la dinámica cuántica diferente de la de Schrödinger%
\footnote{Bajo condiciones muy generales, las dinámicas proporcionadas por ambas
ecuaciones son equivalentes, aunque esto no es cierto con toda generalidad.
La Mecánica Cuántica en la formulación de Born, Heisenberg, Jordan
y Dirac es más general que la de Schrödinger.%
}. Para comprender su significado, recordemos brevemente la formulación
de la dinámica Newtoniana en términos de los corchetes de Poisson.
Suponiendo una partícula que se desplaza en $\mathbb{R}^3$ siguiendo la curva
$\mathbf{x}=\mathbf{x}(t)$
bajo la influencia de un potencial $V\in\mathcal{C}^{\infty}(U)$
(con $U\subset\mathbb{R}^{3}$ una cierta región que podemos suponer
abierta), la ecuación de Newton es
\[
\frac{d^{2}\mathbf{x}}{dt}(t)=-\nabla V\circ\mathbf{x}(t).
\]
Introduciendo la variable $\mathbf{v}(t)=\frac{d\mathbf{x}}{dt}$
y llamando $\tilde{x}(t)=(\mathbf{x}(t),\mathbf{v}(t))$ a la curva
en $U\times\mathbb{R}^{3}$ determinada por $\mathbf{x}(t)$ y su
derivada%
\footnote{En realidad, $\tilde{x}(t)$ es el levantamiento de la curva $\mathbf{x}(t)$
al fibrado cotangente $T^{*}U=\bigsqcup_{\mathbf{x}\in U}T_{\mathbf{x}}^{*}U$,
donde $\mathbf{v}(t)\in T_{\mathbf{x}(t)}U$ se identifica con su
imagen mediante el isomorfismo $T_{\mathbf{x}(t)}^{*}U\simeq T_{\mathbf{x}(t)}U$
inducido por la métrica asociada a la energía cinética. Pero para
el propósito de estas notas, podemos ignorar estas sutilezas.%
}, el sistema anterior se transforma en:
\begin{equation}
\begin{cases}
\mathbf{v}(t)=\frac{d\mathbf{x}}{dt}(t)\\
\frac{d\mathbf{v}}{dt}(t)=-\nabla V\circ\mathbf{x}(t)
\end{cases}.\label{eq_hamilton1}
\end{equation}
En términos del Hamiltoniano del sistema, que es la función $H\in\mathcal{C}^{\infty}(U\times\mathbb{R}^{3})$
dada por\[
H(\mathbf{x},\mathbf{v})=\frac{1}{2}||\mathbf{v}||^{2}+V(\mathbf{x}),\]
las ecuaciones (\ref{eq_hamilton1}) se escriben (para $i\in\{1,2,3\}$)
\begin{equation}
\begin{cases}
\frac{dx^{i}}{dt}(t)=\frac{\partial H}{\partial v^{i}}\circ\tilde{x}(t)\\
\frac{dv^{i}}{dt}(t)=\frac{\partial H}{\partial x^{i}}\circ\tilde{x}(t)
\end{cases},\label{eq_hamilton2}
\end{equation}
que son conocidas como ecuaciones de Hamilton. Para resolver estas
ecuaciones, que son equivalentes a las de Newton para el caso que
estamos considerando, es preciso dar unas condiciones iniciales sobre
$\mathbf{x}(t)$ y $\mathbf{v}(t)$. En otras palabras, para determinar
el estado futuro de un sistema clásico es necesario tener las ecuaciones
de evolución (\ref{eq_hamilton2}) y conocer $\mathbf{x}(t)$, $\mathbf{v}(t)$
en un instante dado $t_{0}$. Por eso, en Mecánica Clásica a $U\times\mathbb{R}^{3}$
se le denomina espacio de estados del sistema (a veces, también se
le denomina espacio de fases, aunque no son exactamente lo mismo,
pues el espacio de fases es un subfibrado del cotangente, cfr. la
nota a pie de página precedente).

Supongamos ahora que queremos evaluar una función $f$ a lo largo
de una trayectoria $\tilde{x}(t)=(\mathbf{x}(t),\mathbf{v}(t))$ en
el espacio $U\times\mathbb{R}^{3}$, donde introducimos coordenadas
$\{x^{1},x^{2},x^{3},v^{1},v^{2},v^{3}\}$. Se tiene que, aplicando
la regla de la cadena y las ecuaciones de Hamilton,
\begin{equation}
\begin{array}{ccl}
\frac{d(f\circ\tilde{x})}{dt}(t) & = & \left.\frac{\partial f}{\partial x^{i}}\right|_{\tilde{x}(t)}\frac{dx^{i}}{dt}(t)+\left.\frac{\partial f}{\partial v^{j}}\right|_{\tilde{x}(t)}\frac{dv^{j}}{dt}(t)\\
\\ & = & \sum_{i=1}^{3}\left.\frac{\partial f}{\partial x^{i}}\right|_{\tilde{x}(t)}\frac{\partial H}{\partial v^{i}}(\tilde{x}(t))-\left.\frac{\partial f}{\partial v^{i}}\right|_{\tilde{x}(t)}\frac{\partial H}{\partial x^{i}}(\tilde{x}(t))\\
\\ & = & \sum_{i=1}^{3}\left(\frac{\partial f}{\partial x^{i}}\frac{\partial H}{\partial v^{i}}-\frac{\partial f}{\partial v^{i}}\frac{\partial H}{\partial x^{i}}\right)\circ\tilde{x}(t)
\end{array}.
\label{eq_hamilton3}
\end{equation}
Esto se suele escribir en forma más compacta omitiendo la trayectoria
particular $\tilde{x}(t)$ e introduciendo el llamado corchete de
Poisson, una aplicación $\mathbb{R}-$bilineal 
$\{.,.\}:\mathcal{C}^{\infty}(U\times\mathbb{R}^{3})\times
\mathcal{C}^{\infty}(U\times\mathbb{R}^{3})
\rightarrow\mathcal{C}^{\infty}(U\times\mathbb{R}^{3})$
definida por
\begin{equation}
\{f,g\}=\sum_{i=1}^{3}\left(\frac{\partial f}{\partial x^{i}}\frac{\partial g}{\partial v^{i}}-\frac{\partial f}{\partial v^{i}}\frac{\partial g}{\partial x^{i}}\right).
\label{eq_poisson}
\end{equation}
Las ecuaciones (\ref{eq_hamilton3}) resultan ser, entonces,
\begin{equation}
\frac{df}{dt}=\{f,H\},\label{eq_poisson2}
\end{equation}
llamada la ecuación de Poisson. Lo que hizo Dirac fue postular una
ecuación dinámica cuántica análoga a (\ref{eq_poisson2}), pero reemplazando
los observables clásicos (en general funciones de $ \mathcal{C}^{\infty}(U\times\mathbb{R}^{n}) $)
por observables cuánticos (operadores de $End_\mathbb{C} (L^{2}(\mathbb{R}^n))$).
Para el caso $n=1$, esto se traduce en las llamadas ``reglas de cuantización canónicas''
\begin{eqnarray*}
x \mapsto x \cdot \\ 
p \mapsto -i\hbar \frac{d}{dx}\cdot
\end{eqnarray*}
y el corchete de Poisson por el llamado conmutador cuántico $[.,.]$ definido
mediante:
\[
[A,B]=\frac{i}{\hbar}\left( A\circ B-B\circ A \right),\mbox{ }\forall A,B\in End_{\mathbb{C}}(L^2 (\mathbb{R}))
\]
(véase el Apéndice B, donde se reproduce el razonamiento de Dirac
para llegar a esta expresión) de manera que si $A\in End_{\mathbb{C}}(L^2 (\mathbb{R}))$
es un observable cualquiera que depende de posiciones, velocidades e
implícitamente del tiempo (pero \emph{no} explícitamente) su evolución está
dada por la \emph{ecuación de Dirac-Heisenberg}:
\[
\frac{dA}{dt}=[H,A],
\]
siendo $H\in End_{\mathbb{C}}(L^2 (\mathbb{R}))$ el Hamiltoniano cuántico del sistema.

\section{El teorema de conexión espín-estadística}

Mediante experimentos del tipo Stern-Gerlach, pronto se pudo determinar
que las part\'{\i}culas elementales se pod\'{\i}an clasificar según
su valor del espín en dos grupos: aquéllas con espín entero, llamadas
\emph{bosones}, y aquellas con espín semientero, llamadas \emph{fermiones}.
Ejemplos de la primera clase son los fotones (espín $1$), y de la
segunda, como ya hemos mencionado, los electrones (espín $1/2$). La
denominación ``bosón'' y ``fermión'' se debe a la diferente
estad\'{\i}stica que obedecen estas part\'{\i}culas: de Bose-Einstein
o de Fermi-Dirac, respectivamente. Con el fin de explicar esta frase,
supongamos ahora que disponemos de un conjunto de part\'{\i}culas
indistinguibles, tales como los electrones en un haz de átomos de
plata, aunque para simplificar la discusión sólo consideraremos dos
de ellas. La propiedad de indistinguibilidad se traduce en que
\begin{equation}
H(1,2)=H(2,1),\label{eq0_1}
\end{equation}
 donde ``$1$'' representa todo el conjunto de parámetros (llamados 
 ``números cuánticos'' en Física)
necesarios para describir la part\'{\i}cula $1$ (inclu\'{\i}do su
espín), y lo mismo ``$2$'' para la part\'{\i}cula $2$. La ecuación
de Schrödinger para este sistema de dos part\'{\i}culas se escribe
entonces como $H(1,2)\Psi(1,2)=E\Psi(1,2)$, aunque al no importar
cómo etiquetemos a las part\'{\i}culas, esto es equivalente a 
$H(2,1)\Psi(2,1)=E\Psi(2,1)$.

Ahora bien, haciendo uso de (\ref{eq0_1}), también resulta
\begin{equation}
H(1,2)\Psi(2,1)=E\Psi(2,1).\label{eq0_2}
\end{equation}

Llegados a este punto, conviene introducir el llamado \emph{operador
de intercambio} $P_{12}$, que aplicado a un estado intercambia todas
las coordenadas (espaciales y espín) de las part\'{\i}culas $1$ y
2$,$ es decir, $P_{12}\Psi(1,2)=\Psi(2,1)$. As\'{\i}, de (\ref{eq0_2})
\begin{equation*}
HP_{12}\Psi(1,2) =E\Psi(2,1)=EP_{12}\Psi(1,2)
=P_{12}E\Psi(1,2)=P_{12}H\Psi(1,2),
 \end{equation*}
 o, lo que es lo mismo en términos del conmutador entre operadores
 \footnote{Sobre la notación: muchos autores denominan $[\_,\_]$ a lo
 que nosotros estamos llamando conmutador de endomorfismos, $[\_,\_]\_$.
 Estos autores escriben la ecuación de Dirac-Heisenberg como
 \begin{equation*}
i\hbar\frac{dA}{dt}=[A,H].
 \end{equation*}},
 definido por $[A,B]\_ =A\circ B-B\circ A$,
\begin{equation}
[H,P_{12}]\_ =0.\label{eq0_3}
\end{equation}

Notemos que, como hemos señalado en la sección precedente, la estructura
formal de la Mecánica Cuántica en la forma en que la estamos utilizando
(à la Heisenberg), es idéntica a la de la Mecánica Clásica en la formulación
de Poisson reeemplazando el corchete clásico de Poisson $\{\_,\_\}$
por el conmutador cuántico $[\_,\_]$. En particular, la evolución temporal
de las magnitudes representadas por los operadores como $P_{12}$
viene dada por su conmutador con el Hamiltoniano, por lo que (\ref{eq0_3})
nos dice que el operador de intercambio es una constante del movimiento.
Otra propiedad importante de este operador es su idempotencia: $(P_{12})^{2}\Psi(1,2)=\Psi(1,2)$,
as\'{\i} que sus valores propios son $\pm1$; sus funciones propias
son las combinaciones simétricas y antisimétricas 
\begin{equation}
\begin{array}{l}
\Psi^{S}(1,2)=\frac{1}{\sqrt{2}}\left(\Psi(1,2)+\Psi(2,1)\right)\\
\text{y}\\
\Psi^{A}(1,2)=\frac{1}{\sqrt{2}}\left(\Psi(1,2)-\Psi(2,1)\right)\text{.}
\end{array}\label{eq0_4}
\end{equation}

El que $P_{12}$ sea una constante del movimiento, implica que un
estado que es simétrico en un instante inicial siempre será simétrico
y que un estado antisimétrico siempre seguirá siendo antisimétrico.
Lo verdaderamente importante (tanto que a su descubridor, Pauli, le
valió el Nobel en 1945), es que \emph{la simetr\'{\i}a o antisimetr\'{\i}a
bajo el intercambio de dos part\'{\i}culas es una caracter\'{\i}stica
de las part\'{\i}culas}, y no algo que se pueda decidir en la preparación
del estado inicial. Esta propiedad se conoce como teorema de conexión
espín-estad\'{\i}stica%
\footnote{A veces se habla del ``principio\textquotedblright\ de conexión
espín-estad\'{\i}stica, pero hay que hacer notar que uno de los logros
(quizás \emph{el} logro) de la teor\'{\i}a cuántica de campos axiomática,
es que este enunciado se deriva como un teorema
a partir de unos postulados, véase \cite{Str-Wig 89}.} y afirma que:

\begin{enumerate}
\item Los sistemas consistentes en part\'{\i}culas idénticas con \emph{espín
semientero} se describen mediante funciones de onda antisimétricas,
y se dice que obedecen la \emph{estad\'{\i}stica de Fermi-Dirac}. 
\item Los sistemas consistentes en part\'{\i}culas idénticas con \emph{espín
entero} se describen mediante funciones de onda simétricas, y se dice
que obedecen la \emph{estad\'{\i}stica de Bose-Einstein}. 
\end{enumerate}

\section{Principio de exclusión y números de ocupación}

Una consecuencia inmediata del teorema de conexión espín-estad\'{\i}stica
que \emph{en un sistema no puede haber más de dos fermiones en un
estado de energ\'{\i}a, momento angular, paridad, etc... definidos}%
\footnote{En un hipotético sistema de tres o más fermiones, dos de ellos tendr\'{\i}an
la variable de espín con valores opuestos y el tercero repetiría los
valores de esta variable, lo cual, por la antisimetría de la función
de onda, implicaría que ésta es nula%
}, limitación que no existe en el caso de los bosones. Este enunciado,
se conoce como \emph{Principio de exclusión de Pauli}. Vamos a ver
con algo de detalle esta implicación, ya que es la clave para la introducción
de las ideas de la Supersimetr\'{\i}a (al menos por lo que respecta
a estas notas).

Supongamos pues un sistema de $N$ fermiones \emph{indistinguibles}
(digamos $N$ electrones), descritos por las funciones de onda respectivas
$\psi_{1}(\nu_{1}),...,\psi_{N}(\nu_{N})$ (pensemos que en $\nu$
van inclu\'{\i}dos todos los números cuánticos del sistema, pero esto
son sólo etiquetas: ¡las part\'{\i}culas son indistinguibles!). Recordemos
de la sección \ref{estados} que esto significa, en el ejemplo de
part\'{\i}culas de espín $\frac{1}{2}$ (recuérdese (\ref{eq0_04a}))\[
\psi_{i}(\nu_{i})=\alpha_{+}\left\vert t,\mathbf{x},+\right\rangle +\alpha_{-}\left\vert t,\mathbf{x},-\right\rangle .\]
 El intento naïf de formar una función de onda común a partir de las
de cada part\'{\i}cula es, obviamente, el producto tensorial $\psi_{1}(\nu_{1})\otimes\cdots\otimes\psi_{N}(\nu_{N})$,
pero lamentablemente esta función no satisface el requisito de antisimetr\'{\i}a.
Una manera fácil de conseguir que si se cumpla, es mediante el operador
de antisimetrización%
\footnote{Fijémonos en que este operador no es otro que el operador $Alt$ que
pasa del producto tensorial al exterior sobre módulos o espacios vectoriales.%
}: se toma la función $\Psi$ definida por \[
\Psi=\frac{1}{\sqrt{N!}}\sum\limits _{\sigma\in S_{N}}sig(\sigma)\psi_{1}(\nu_{\sigma(1)})\otimes\cdots\otimes\psi_{N}(\nu_{\sigma(N)})\]
 donde $S_{N}$ es el grupo de permutaciones de $N$ elementos y $sig\left(\sigma\right)$
es la signatura de la permutación $\sigma$. Es claro que $\Psi$
es antisimétrica bajo el intercambio de las coordenadas de dos part\'{\i}culas
cualesquiera (¡es la generalización de (\ref{eq0_4})!). De hecho,
$\Psi$ se puede escribir como un determinante, el llamado \emph{determinante
de Slater} (donde el producto de los factores debe entenderse como
producto tensorial): 
\[
\Psi=\left\vert 
\begin{array}{ccccc}
\psi_{1}(\nu_{1}) & \psi_{1}(\nu_{2}) & \cdots & \psi_{1}(\nu_{N-1}) & \psi_{1}(\nu_{N})\\
\psi_{2}(\nu_{1}) & \psi_{2}(\nu_{2}) & \cdots & \psi_{2}(\nu_{N-1}) & \psi_{2}(\nu_{N})\\
\cdot & \cdot & \cdot & \cdot & \cdot\\
\psi_{N-1}(\nu_{1}) & \psi_{N-1}(\nu_{2}) & \cdots & \psi_{N-1}(\nu_{N-1}) & \psi_{N-1}(\nu_{N})\\
\psi_{N}(\nu_{1}) & \psi_{N}(\nu_{2}) & \cdots & \psi_{N}(\nu_{N-1}) & \psi_{N}(\nu_{N})
\end{array}\right\vert ,
\]
 de modo que es manifiesto que la función de onda total $\Psi$ se
anula idénticamente si dos cualesquiera de las funciones $\psi_{i}(\nu_{j})$
son la misma, es decir: en un sistema de $N$ fermiones no puede haber
dos de ellos en el mismo estado cuántico%
\footnote{Esto sólo es cierto para fermiones en un mismo sistema. No se aplica
a fermiones cuyas funciones de onda están completamente incorrelacionadas.%
}.

Naturalmente, este proceso no agota todas las posibilidades de obtener
funciones de onda totales para un sistema de fermiones que sean antisimétricas.
Por ejemplo, podemos tomar combinaciones lineales de determinantes
de este tipo y el resultado es automáticamente antisimétrico. Esta
idea es la base de una construcción más general llamada en F\'{\i}sica
\emph{espacio de Fock}. Veamos cómo se describe en F\'{\i}sica: nuestras
funciones de onda pertenecen a un cierto espacio de Hilbert separable
que, como buen espacio vectorial provisto de producto escalar complejo
(¡y con respecto a la métrica inducida por el cual es completo!) posee
una base numerable y ortonormal de funciones, $\left\{ \phi_{j}\right\} _{j\in\mathbb{N}}$.
Los estados de una part\'{\i}cula de nuestro sistema se pueden expresar
como el conjunto de límites de todas las combinaciones lineales de las $\phi_{j}\left(\nu\right)$
(tomadas de una en una, es decir, cosas del tipo $\alpha_{1}\phi_{1}+\cdots+\alpha_{k}\phi_{k}+\cdots$),
y se puede probar que los estados antisimétricos de $N$ part\'{\i}culas
están formados por todas las combinaciones lineales de los determinantes
de Slater obtenidos tomando $N$ funciones de entre las $\left\{ \phi_{j}\right\} _{j\in\mathbb{N}}$.
Éste es el llamado espacio de Fock de orden $N$ del sistema%
\footnote{Obsérvese que estamos suponiendo $N$ part\'{\i}culas \emph{indistinguibles}.
Cuando no es as\'{\i}, la estructura del espacio de estados total
si es la del producto tensorial de los espacios individuales de cada
part\'{\i}cula.%
}, y el espacio de Fock total es la clausura de la suma directa de todos ellos para
los diferentes valores de $N$. En realidad, como vemos, el espacio
de Fock es el resultado de tomar la clausura del cociente del álgebra tensorial
del espacio de Hilbert (considerado como simple espacio vectorial)
por el ideal engendrado por la antisimetrización aplicada a los elementos
generadores, es decir: desde un punto de vista matemático no es otra
cosa que la clausura del álgebra exterior del espacio de Hilbert original.

Vamos a introducir una notación más cómoda para estas funciones de
onda antisimétricas de $N$ fermiones, basada en el hecho de que cualquier
espacio de Hilbert separable de dimensión infinita es isomorfo a un
espacio de sucesiones (véase \cite{Von 32}). Supongamos que los
diferentes estados \emph{base} que puede tener \emph{una sola} partícula
se enumeran como
\[
\{\psi_{1}(\nu_{1}),\psi_{2}(\nu_{2}),\psi_{3}(\nu_{3}),...\}=\{\psi_{i}(\nu_{i})\}_{i\in\mathbb{N}};
\]
como ya hemos mencionado, el Principio de exclusión de Pauli hace
que en cada uno de los estados descritos por $\psi_{1}(\nu_{1}),...,$
$\psi_{N}(\nu_{N})$ haya una o ninguna particula, pero nunca dos
o más. Una manera muy conveniente de escribir las funciones, pues,
es indicar los estados que están {}``ocupados'' : aquellos en los
que se encuentra alguna part\'{\i}cula. De este modo, en general escribiremos
\[
\left\vert n_{1},n_{2},...\right\rangle 
\]
 para la función de onda total que describe a $n_{1}$ part\'{\i}culas
en el estado $\psi_{1}$, $n_{2}$ part\'{\i}culas en el estado $\psi_{2}$,
etc. La particularidad es que las $n_{j}$ sólo pueden tomar los valores
$1$ y el resto se sobreentiende que es $0$ (hay una representación
similar para los estados de bosones, pero estos no tienen esa limitación:
$n_{j}$ puede ser cualquier natural). Por ejemplo,
\[
\begin{array}{l}
\left\vert 1_{1}\right\rangle =\left\vert 1,0,0,0,0,0,...\right\rangle =\psi_{1}(\nu_{1})\\
\left\vert 1_{1},1_{3}\right\rangle =\left\vert 1,0,1,0,0,0,...\right\rangle =\frac{1}{2}\left(\psi_{1}(\nu_{1})\otimes\psi_{3}(\nu_{3})-\psi_{3}(\nu_{3})\otimes\psi_{1}(\nu_{1})\right).
\end{array}
\]
En cualquiera de los casos, $n_{1}+n_{2}+\cdots=N$. Esta manera de
describir los estados del sistema, con mucha propiedad se llama \emph{representación
por el número de ocupación}.

\section{Ejemplo: un bosón y un fermión en una caja\label{S8}}

Consideremos un bosón (digamos, un pión $\pi^{0}$ cuya carga y espín
valen $0$) y un fermión (un $e^{-}$, con carga $-1$ y espín $1/2$),
sin interacción entre ellos, moviéndose a lo largo de una recta y
confinados a la región $\ \left]0,L\right[$. La descripción matemática
de esta situación se basa en tomar un potencial $\mathcal{C}^\infty$ a trozos
(llamado de ``pozo cuadrado infinito'' por razones obvias) de la forma
\[
V(x)=\left\{ 
\begin{array}{l}
\infty\text{ si }x\geq L\text{ \'{o} }x\leq0\\
\\
0\text{ si }0<x<L,
\end{array}\right.
\]
 que separa la recta en tres regiones: $\left]-\infty,0\right],\left]0,L\right[,\left[L,+\infty\right[$.
Sólo la segunda de ellas tendrá interés desde el punto de vista del
cálculo (¡y desde el f\'{\i}sico también!), pues obviamente la función
de onda de cualquiera de las part\'{\i}culas en las otras dos es $0$.

Olvidémonos por un momento del espín de las part\'{\i}culas.
Busquemos soluciones separables a la ecuación de Schrödinger para $V(x)$,
esto es, en la forma $\psi(t,x)=\phi(x)\cdot f(t)$. Por lo
que respecta a su comportamiento espacial, cualquiera de ellas vendrá
descrita por una $\phi(x)\in L^{2}(\left]0,L\right[)$ que satisfaga
la ecuación de Schrödinger en la región $\left]0,L\right[\subset\mathbb{R}$
(en la cual $V(x)=0$, cfr. (\ref{eq0_04b}), (\ref{eq0_05})):
\begin{equation}
-\frac{\hbar^{2}}{2m}\phi^{\prime\prime}(x)=E\phi(x),\label{eq0_4_1}
\end{equation}
donde suponemos que $E$ ---la energía de la partícula--- es positiva, en 
tanto que la parte temporal está descrita por una $f(t)\in L^{2}(\left]0,L\right[)$
con\[
f^{\prime}(t)=-\frac{iE}{\hbar}f(t),\]
 o sea\[
f(t)=\exp(-iEt/\hbar),\]
 de modo que la función de onda espacio-temporal ser\'{\i}a\[
\psi(t,x)=\phi(x)\exp(-iEt/\hbar).\]

Ocupémonos ahora de las soluciones a (\ref{eq0_4_1}). El polinomio
caracter\'{\i}stico de la ecuación es $P(\lambda)=\frac{\hbar^{2}}{2m}\lambda^{2}+E$,
con ra\'{\i}ces imaginarias $\lambda=\pm i\sqrt{2mE}/\hbar$, por
lo que una base del espacio de soluciones es la trigonométrica y existirán
unas constantes $A,B$ tales que\[
\phi(x)=A\cos\left(x\sqrt{2mE}/\hbar\right)+B\sin\left(x\sqrt{2mE}/\hbar\right).\]
 En definitiva, nuestra función de onda espacio-temporal para cualquiera
de las part\'{\i}culas en la caja tendrá que ser de la forma \[
\psi(t,x)=\left\{ \begin{array}{l}
\left(A\cos\left(\frac{x\sqrt{2mE}}{\hbar}\right)+B\sin\left(\frac{x\sqrt{2mE}}{\hbar}\right)\right)e^{-iEt/\hbar},\text{ }(x,t)\in\left]0,L\right[\times\mathbb{R}\\
\\0,\text{ }x\geq L\text{ \'{o} }x\leq0.\end{array}\right.\]
 Las constantes $A,B$ se pueden determinar por las condiciones de
frontera, provenientes del hecho de que el potencial está definido por secciones pero 
queremos que las soluciones correspondientes a cada sección se empalmen
adecuadamente para dar funciones $\mathcal{C}^\infty$ (en inglés, a estas condiciones 
se las denomina ``matching conditions''):
\[
\psi(t,0)=0=\psi(t,L),\text{ }\forall t\in\mathbb{R},
\]
 de donde es inmediato que\[
A=0\text{ y }B\sin\left(L\sqrt{2mE}/\hbar\right)=0.\]
 La segunda condición no implica que la constante $E$ (en principio
arbitraria) sea nula, sino que $L\sqrt{2mE}/\hbar=n\pi$ con $n\in\mathbb{Z}-{0}$,
esto es
\[
E=\frac{2n^{2}}{m}\left(\frac{\hbar\pi}{L}\right)^{2},\label{energias}
\]
 de manera que las posibles energ\'{\i}as de las part\'{\i}culas no
pueden tomar valores arbitrarios, sino sólo algunos de ellos parametrizados
por $n\in\mathbb{Z}-\{ 0\}$: ¡ésta es la famosa \emph{cuantización de la
energ\'{\i}a}!. Fijémonos en que, por aparecer la $n$ elevada al cuadrado, podemos
tomar\footnote{Observemos que el valor $n=0$ no es admisible físicamente,
pues conduce a una función de onda nula y ésta no puede cumplir el requisito de
normalización (\ref{eq0_02})} $n \in\mathbb{N}$. Escribiremos $E_{n}$ cuando queramos hacer explícito el
valor concreto a que nos referimos.

Con esto, la función de onda pasa a ser 
\[
\psi(t,x)=\left\{ \begin{array}{l}
B\sin\left(\frac{n\pi}{L}x\right)\exp\left(-i\frac{\hbar^{2} n^{2}}{2m}\left(\frac{\pi}{L}\right)^{2}t\right),\text{ }(x,t)\in\left]0,L\right[\times\mathbb{R}\\
\\0,\text{ }x\geq L\text{ \'{o} }x\leq0.\end{array}\right.
\]
 Para determinar la constante $B$, utilizaremos el hecho de que la
función de onda del sistema debe estar normalizada a $1$:
\[
\int\nolimits _{\mathbb{R}}\psi(t,x)\psi^{\ast}(t,x)dx=1,\text{ }\forall t\in\mathbb{R.}
\]
 En nuestro caso, la ecuación precedente se reduce a
 \[
\int\nolimits _{0}^{L}B^{2}\sin^{2}(\omega x)dx=1,\text{ donde }\omega=\frac{n\pi}{L},
\]
 esto es
 \[
\left[B^{2}\left(\frac{x}{2}-\frac{1}{4\omega}\sin(2\omega x)\right)\right]_{0}^{L}=1,
\]
 de donde resulta
 \[
B=\sqrt{\frac{2}{L}}e^{i\varphi},
\]
con $\varphi\in [0,2\pi]$. Por tanto, la función de onda espacio-temporal de cualquiera de las
part\'{\i}culas será una de la familia, para $n \in \mathbb{N}$, 
\begin{equation}
\psi_{n}(t,x)=\left\{ \begin{array}{l}
\sqrt{\frac{2}{L}}\sin\left(\frac{n\pi}{L}x\right)\exp\left(-i\left(
\frac{\hbar^{2} n^{2}}{2m}\left(\frac{\pi}{L}\right)^{2}t-\varphi\right)\right),\text{ }(x,t)\in\left]0,L\right[\times\mathbb{R}\\
\\0,\text{ }x\geq L\text{ \'{o} }x\leq0.\end{array}\right.\label{fdeonda}
\end{equation}
Nótese la indeterminación de la función de onda $\psi_{n}(t,x)$ introducida
por el factor de fase $e^{i\varphi}$. La notación habitual (en términos de los bra y 
kets de Dirac) es
$\psi_{n}(t,x)=\left\vert t,x,n\right\rangle .$

Para el pión $\pi^{0}$, que tiene espín $0$, éstas son las funciones
de onda totales. Ahora podemos considerar la inclusión del espín para
el caso del fermión, que dará como función de onda total el producto
tensorial de $\psi_{n}(t,x)$ por una de las funciones espinoriales
$\zeta=\alpha\eta_{\uparrow}+\beta\eta_{\downarrow}$ que ya hemos
estudiado. Con mayor precisión, para el $\pi^{0}$ tenemos que el
único número cuántico es $n$ ---esto es, $\nu$ es un conjunto de parámetros
con un único elemento: $n$--- y los estados base serán $\{\psi_{i}(\nu_{i})\}_{i\in\mathbb{N}},$
donde para cada $i\in\mathbb{N},$ $\psi_{i}(t,x)$ está dada por
(\ref{fdeonda}). Abreviadamente, se suelen denotar estas funciones
de onda por $\{\left\vert i\right\rangle \}_{i\in\mathbb{N}}$. Para
el fermión, podemos numerar los estados base, de acuerdo con (\ref{eq0_04a}),
como
\[
\{\left\vert 1,+\right\rangle ,\left\vert 1,-\right\rangle ,\left\vert 2,+\right\rangle ,\left\vert 2,-\right\rangle ,...\},
\]
donde, recordemos, $\left\vert i,\pm\right\rangle =\left\vert \psi_{i}(t,x)\right\rangle \otimes\left\vert \pm\right\rangle $.

\section{El álgebra de los operadores de creación y aniquilación}

Volvamos al estudio de un conjunto de $N$ part\'{\i}culas fermiónicas
indistinguibles. Nos interesa ahora una situación más dinámica, en
la que los fermiones de nuestro sistema pueden pasar de un estado
a otro. Esta es la situación que uno estudia, por ejemplo, en teor\'{\i}a
cuántica de campos (QFT), donde se va incluso un paso más allá y se
admite no sólo que los estados de las diferentes part\'{\i}culas puedan
cambiar, sino también el propio número de éstas. Salvo algunas particularidades
técnicas (¡muy importantes!) las ideas centrales de la QFT en su presentación
estándar se basan en una generalización del formalismo que vamos a
introducir ahora para describir cambios de estado, a uno que describa
procesos de creación y aniquilación de las propias part\'{\i}culas.

Definimos, formalmente, un \emph{operador creación} que crea un estado de una part\'{\i}cula
en el modo $k-$simo, $b_{k}^{\dag}$: 
\[
b_{k}^{\dag}\left\vert n_{1},n_{2},...,n_{k},...\right\rangle =\left\vert n_{1},n_{2},...,n_{k}+1,...\right\rangle ,
\]
 y un \emph{operador aniquilación} que destruye un modo, $b_{k}$:
\[
b_{k}\left\vert n_{1},n_{2},...,n_{k},...\right\rangle =\left\vert n_{1},n_{2},...,n_{k}-1,...\right\rangle .
\]
 Naturalmente, hay que establecer alguna regla adicional para estos
operadores, pues no podemos violar el Principio de exclusión y tener
más de una part\'{\i}cula en el mismo estado. Como además sabemos
(por la misma razón) que los $n_{j}$ de partida no pueden valer más
de $1$, tenemos las reglas obvias:
\[
(b_{k}^{\dag})^{2}=0,\text{ y }(b_{k})^{2}=0,
\]
 actuando sobre cualquier estado. As\'{\i}, tendremos, por ejemplo
(si $\left\vert 0\right\rangle $ representa el estado en que todos
los modos del sistema están desocupados):
\[
b_{k}^{\dag}\left\vert 0\right\rangle =\left\vert 1_{k}\right\rangle =\psi_{k}(\nu)
\]
 y también\footnote{Recuérdese que el orden de actuación de los operadores
(que se traduce en el orden en que se toman los términos del producto tensorial)
es importante.}, 
 \[
b_{k^{\prime}}^{\dag}b_{k}^{\dag}\left\vert 0\right\rangle =\left\vert 1_{k^{\prime}},1_{k}\right\rangle =\frac{1}{2}\left(\psi_{k}(\nu_{k})\otimes\psi_{k^{\prime}}(\nu_{k^{\prime}})-\psi_{k^{\prime}}(\nu_{k^{\prime}})\otimes\psi_{k}(\nu_{k})\right).
\]
 Fijémonos en que la antisimetr\'{\i}a de la función de onda total
(i.e: $\left\vert 1_{k^{\prime}},1_{k}\right\rangle =-\left\vert 1_{k},1_{k^{\prime}}\right\rangle$ implica que 
\begin{equation}
b_{k^{\prime}}^{\dag}b_{k}^{\dag}\left\vert 0\right\rangle =-b_{k}^{\dag}b_{k^{\prime}}^{\dag}\left\vert 0\right\rangle ,\label{eq0_4b}
\end{equation}
 es decir: 
 \[
(b_{k}^{\dag}b_{k^{\prime}}^{\dag}+b_{k^{\prime}}^{\dag}b_{k}^{\dag})\left\vert 0\right\rangle =0.
\]
 No hay nada de especial en este cálculo que se refiera al $\left\vert 0\right\rangle $,
de hecho, podemos escribir con generalidad
\[
b_{k}^{\dag}b_{k^{\prime}}^{\dag}+b_{k^{\prime}}^{\dag}b_{k}^{\dag}=0\text{,}
\]
 y es costumbre abreviar esta expresión escribiéndola en términos
del anticonmutador $[ .,.]_{+} $, definido para un par de
operadores $A$, $B$ como $[A,B]_{+}=A\circ B+B\circ A $:
\[
[b_{k}^{\dag},b_{k^{\prime}}^{\dag}]_{+}=0.
\]
 As\'{\i}, decimos que \emph{los operadores de creación anticonmutan
entre ellos} (observemos que el caso $k=k^{\prime}$ también está
trivialmente inclu\'{\i}do). De manera totalmente análoga se ve que
también \emph{los operadores de aniquilación anticonmutan entre ellos}:
\[
[b_{k},b_{k^{\prime}}]_{+} =0,
\]
 as\'{\i} que para establecer el álgebra de los operadores de creación
y aniquilación, debemos estudiar qué ocurre con su anticonmutador
cuando cada uno es de un tipo. Es trivial que si $k\neq k^{\prime}$
se tiene:
\[
[b_{k}^{\dag},b_{k^{\prime}}]_{+}=0,
\]
 luego podemos centrar nuestra atención en el caso de $[b_{k}^{\dag},b_{k}]_{+}$.

Consideremos las siguientes actuaciones, que se deducen de las definiciones
y del Principio de exclusión de Pauli:
\begin{equation}
\left\{ \begin{array}{ll}
\begin{array}{l}
b_{k}^{\dag}b_{k}\left\vert 0_{k}\right\rangle =0,\end{array} & \begin{array}{l}
b_{k}b_{k}^{\dag}\left\vert 0_{k}\right\rangle =\left\vert 0_{k}\right\rangle \end{array}\\
b_{k}^{\dag}b_{k}\left\vert 1_{k}\right\rangle =\left\vert 1_{k}\right\rangle , & b_{k}b_{k}^{\dag}\left\vert 1_{k}\right\rangle =0;
\end{array}\right.\label{eq0_5}
\end{equation}
 sumándolas con coeficientes arbitrarios $\alpha$ y $\beta$, resulta:
\[
\alpha(b_{k}^{\dag}b_{k}+b_{k}b_{k}^{\dag})\left\vert 0_{k}\right\rangle +\beta\left\vert 1_{k}\right\rangle =\alpha\left\vert 0_{k}\right\rangle +\beta\left\vert 1_{k}\right\rangle ,
\]
as\'{\i} que para operaciones efectuadas sobre cualquier vector que
describa el modo $k-$simo,
\[
b_{k}^{\dag}b_{k}+b_{k}b_{k}^{\dag}=1.
\]
 Por supuesto, el modo $k-$simo es totalmente arbitrario, es decir,
realmente tenemos un resultado general:
\[
[b_{k}^{\dag},b_{k}]_{+}=1.
\]
 De forma unificada, podemos resumir las fórmulas obtenidas diciendo
que
\begin{equation}
[b_{k}^{\dag},b_{k^{\prime}}]_{+}=\delta_{kk^{\prime}}.\label{eq0_6}
\end{equation}
 Además, de (\ref{eq0_5}) resulta evidente que el operador $n_{k}=b_{k}^{\dag}b_{k}$
juega el papel de \emph{operador número de ocupación del estado} $k-$\emph{simo},
y de hecho as\'{\i} se le conoce. Estos operadores son muy importantes
en el esquema algebraico de la Mecánica Cuántica; de hecho, es bien
sabido que definiendo el \emph{operador número total} $N=\sum\limits _{k}n_{k}$,
el Hamiltoniano del oscilador armónico cuántico (bosónico) se expresa
como
\[
H=N+\frac{1}{2}
\]
(véase, por ejemplo, \cite{Gas 03} para profundizar en este aspecto).

Hasta ahora hemos trabajado con fermiones, pero el mismo análisis formal
puede llevarse a cabo con bosones, part\'{\i}culas para las cuales
no existe la restricción impuesta por el Principio de exclusión de
Pauli. Revisando, o mejor dicho, repitiendo el argumento que hemos
dado y haciendo los cambios pertinentes%
\footnote{Ejercicio para el lector%
} (especialmente en (\ref{eq0_4b}), ¡donde ahora hay que tener en
cuenta la \emph{simetr\'{\i}a}!), es fácil convencerse de que para
el caso de los bosones, el álgebra de los operadores de creación-aniquilación,
que ahora denotaremos respectivamente por $a_{k}^{\dag},a_{k^{\prime}}$,
viene dada por las relaciones
\begin{equation}
[a_{k}^{\dag},a_{k^{\prime}}]\_=\delta_{kk^{\prime}},\label{eq0_7}
\end{equation}
donde $[\_,\_]\_$ es el \emph{conmutador} entre dos operadores, que,
recordemos, puede definirse sobre los endomorfismos de cualquier espacio vectorial
(en particular sobre los operadores de nuestro espacio de Hilbert-Fock):
si $V$ es un espacio vectorial y $S,T\in\mathrm{End}V,$
\[
[S,T]\_=S\circ T-T\circ S.
\]

El resto de estas notas se dedicará a dos cosas:
\begin{enumerate}
\item Formalizar estas construcciones en el contexto de las superálgebras de Lie.
\item Estudiar algunas consecuencias particulares, como la posibilidad de dar una
realización de una superálgebra particular (la de Heisenberg) y sus aplicaciones
en Física.
\end{enumerate}

\section{Superálgebras de Lie, teor\'{\i}as gauge y supervariedades}

Fijémonos en las ecuaciones (\ref{eq0_6}) y (\ref{eq0_7}): tienen
un parecido muy llamativo, de hecho sólo difieren en un signo, que
distingue al conmutador del anticonmutador, ¡seguro que hay alguna
manera de escribirlas de forma unificada!. Esta pudo ser la pregunta
(o el reto) que se planteó Bertram Kostant a mediados de los años
70, cuando introdujo la noción de supervariedad a partir de las superálgebras
de Lie. Vamos a ver qué son esas álgebras y qué tienen que ver con
el problema que hemos venido considerando hasta ahora, el de describir
un sistema de part\'{\i}culas idénticas, bosones y fermiones.

La idea intuitiva para unificar el tratamiento de (\ref{eq0_6}) y
(\ref{eq0_7}) puede expresarse as\'{\i}: si convenimos en asignar
a los bosones y fermiones una etiqueta, un {}``grado'' que los identifique
y distinga, podr\'{\i}amos decir que asociando a los bosones el $0$
y a los fermiones el $1$, sus operadores de creación-aniquilación
vienen descritos por
\begin{equation}
[\![c_{k}^{\dag},c_{k^{\prime}}]\!]=\delta_{kk^{\prime}},\label{eq0_8}
\end{equation}
 donde $c$ puede ser $a$ o $b$ y $[\![\_,\_]\!]$, el \emph{superconmutador},
está dado por
\begin{equation}
[\![S,T]\!]=S\circ T-(-1)^{grad(S)\cdot grad(T)}T\circ S,\label{eq0_9}
\end{equation}
 siendo $grad(S)$ el grado de $S$ y $grad(T)$ el de $T$. Obviamente,
(\ref{eq0_8}) se reduce a (\ref{eq0_6}) y (\ref{eq0_7}), respectivamente,
cuando $c=a$ y $c=b$.

Llegados a este punto, podr\'{\i}amos pensar lo siguiente: ahora que
ya sabemos que nuestra descripción de los fenómenos asociados a fermiones
y bosones requiere de dos tipos de operadores distintos (unos que
conmutan y otros que anticonmutan), ¿\emph{no podr\'{\i}amos comenzar
desde el principio definiendo algún tipo de espacios en los que se
pueda desarrollar todo el análisis precedente pero que trate por igual
a los bosones y fermiones}?. La respuesta, que debemos fundamentalmente
a Kostant y Berezin (consultar por ejemplo \cite{Kos 77}, \cite{Ber 87}
y las referencias de éste último para un seguimiento histórico) es
que \emph{sí}, y la gu\'{\i}a nos la dan las ecuaciones (\ref{eq0_8})
y (\ref{eq0_9}).

Se trata de construir un álgebra con un producto $\circ$ de manera que
nos describa formalmente la situación que tenemos para los operadores
de creación-aniquilación. Como ya hemos mencionado, toda la construcción se
basa en la introducción de un ``grado'': consideremos un espacio vectorial $V$
sobre un cuerpo $\mathbb{K}$, que se puede expresar como
suma directa de dos espacios $V=V_{0}\oplus V_{1}$. Los elementos
de $V_{0}$ se llaman (vectores) \emph{pares}, y los de $V_{1}$ (vectores)
\emph{impares}. Se dice entonces que sobre $V$ se ha definido una
$\mathbb{Z}_{2}-$\emph{graduación}, y que $V$ es un \emph{espacio
vectorial $\mathbb{Z}_{2}-$graduado} o un \emph{superespacio vectorial}. 
Si $S:V\rightarrow V$ es un
endomorfismo (es decir, una aplicación lineal), puede ocurrir que
lleve los elementos de $V_{0}$ en los de $V_{0}$, o en los de $V_{1}$,
o incluso parte en uno y otro subespacio. Consideremos sólo los que no mezclan elementos,
y llamémoslos \emph{endomorfismos graduados homogéneos}, $\mathrm{End}_{\mathbb{K}}^{G}(V)$.
Tendremos entonces aplicaciones de dos tipos: las pares (que transforman cada $V_{i}$
con $i=0,1$ en sí mismo) y las impares, que intercambian $V_{0}$ con $V_{1}$. A
las primeras, se les asigna grado $0$ y a las segundas, grado $1$.

Ahora, sobre los endomorfismos de un espacio vectorial graduado podemos
definir varias estructuras algebraicas: una de ellas es la de espacio
vectorial (sobre el mismo cuerpo que $V$), mediante la suma de aplicaciones
y el habitual producto por un escalar. Otra es una estructura de semigrupo,
en la que el producto es la composición de aplicaciones $S\circ T$.
Este producto es asociativo y tiene neutro (la aplicación identidad),
pero en general no todo elemento tiene inverso y, desde luego, no
es conmutativo. Con esto, $(\mathrm{End}_{\mathbb{K}}^{G}(V),+,\circ)$ se dice
que tiene una estructura de \emph{álgebra} $\mathbb{Z}_{2}-$\emph{graduada} o
\emph{superálgebra}.

Una medida de la no conmutatividad de $\circ$ la da el llamado \emph{conmutador
de endomorfismos graduados}:
\begin{equation}
[\![S,T]\!]=S\circ T-(-1)^{grad(S)grad(T)}T\circ S\label{eq0_10}
\end{equation}
(fijémonos en que esto no es otra cosa que lo que antes hemos llamado
superconmutador, cuando $V$ es el espacio de Hilbert-Fock). Es claro
que se tienen las siguientes propiedades%
\footnote{Ejercicio para el lector%
}:

\begin{enumerate}
\item \label{1}$\mathbb{K}-$bilinealidad (con $\mathbb{K}$ el cuerpo
base de $V$). 
\item \label{2}$[\![S,T]\!]=-(-1)^{ST}[\![T,S]\!]$ .
\item \label{3}$[\![S,[\![T,U]\!]]\!]=[\![[\![S,T]\!],U]\!]+(-1)^{ST}[\![T,[\![S,U]\!]]\!]$. 
\end{enumerate}
Con estas propiedades, $(\mathrm{End}_{\mathbb{K}}^{G}(V),[\![\_,\_]\!])$ tiene
estructura de lo que se denomina \emph{superálgebra de Lie}. En la sección siguiente
trataremos estos conceptos con mayor detalle, utilizando un ejemplo concreto.

Notemos que estas son generalizaciones de las estructuras con las que uno trabaja
habitualmente, incluso a nivel clásico: los espacios que aparecen
en Mecánica Clásica como espacios de fase o de configuración son variedades
diferenciales $M$ modeladas sobre espacios vectoriales $\mathbb{R}^{n}$,
es decir, espacios que \emph{localmente} son como un abierto de $\mathbb{R}^{n}$
(piénsese en el espacio de configuración de un péndulo, por ejemplo,
que es la circunferencia $S^{1}$ muy distinta globalmente de $\mathbb{R}$),
y los observables clásicos son las funciones de $C^{\infty}(T^{\ast}M)$
($T^{\ast}M$ es el espacio cotangente), equipadas con el corchete
de Poisson $\{\_,\_\}$ que convierte a $(C^{\infty}(M),\{\_,\_\})$
en un álgebra de Lie: sus propiedades son las mismas que las (\ref{1}),
(\ref{2}), (\ref{3}) anteriores
\footnote{En realidad, el corchete de Poisson $\{\_,\_\}$ tiene una importantísima
propiedad adicional: si $f,g,h \in C^{\infty}(M)$, entonces,
$$
\{f,gh\}=\{f,g\}h+g\{f,h\}.
$$
Se dice que $\{\_,\_\}$ actúa sobre $C^{\infty}(M)$ mediante derivaciones y que
$(C^{\infty}(M),\{\_,\_\})$ es un álgebra de Poisson. En estas notas
tendremos necesidad de recurrir a esta propiedad adicional en el apéndice B.},
pero con la graduación trivial (todos los elementos son pares). As\'{\i}, reemplazando
los espacios vectoriales (modelos locales) de la teor\'{\i}a clásica
por sus análogos $\mathbb{Z}_{2}-$graduados, resulta un marco de
trabajo que retiene todas las caracter\'{\i}sticas algebraicas y anal\'{\i}ticas
necesarias para poder hacer mecánica (se tienen las mismas estructuras
que en el caso clásico, pero ahora graduadas) y se obtiene la ventaja
de un tratamiento simétrico (o mejor dicho, supersimétrico) \emph{ab
initio} de bosones y fermiones, según hemos visto.

Los espacios que localmente están modelados sobre productos de abiertos
de $\mathbb{R}^{n}$ y una superálgebra, se denominan
\emph{supervariedades}, y la categor\'{\i}a de tales espacios es en
la que tiene cabida de manera natural las teor\'{\i}as que pretenden
unificar los distintos tipos de part\'{\i}culas (junto con sus interacciones)
que conocemos hoy en dia.

Por ejemplo, pensemos en la descripción de las part\'{\i}culas mediante
campos. Aqu\'{\i} los objetos fundamentales son aplicaciones $\Psi(x^{\mu}),$
definidas en el espacio-tiempo y con valores en algún espacio vectorial
sobre el que actúa un cierto grupo de simetr\'{\i}a $G$, que en F\'{\i}sica
se denomina grupo de gauge de la teor\'{\i}a. Los espacios vectoriales
en que toman valores los distintos campos correspondientes a part\'{\i}culas
de distinto tipo tienen caracter\'{\i}sticas muy diferenciadas, lo
cual hace que las teor\'{\i}as gauge clásicas tengan que trabajar
con fermiones y bosones por separado, cosa que no sucede si se consideran
supervariedades.

Con algo más de detalle técnico, podemos decir que las teor\'{\i}as
gauge clásicas se basan en la consideración de una variedad espacio-tiempo
$M_{4}$, sobre la que se tiene un fibrado principal $(P,\pi,M_{4},G)$;
los objetos f\'{\i}sicos de interés son de dos tipos: los campos de
materia (que describen electrones, muones, etc), que se representan
por secciones de un fibrado vectorial $(E,\tau,M_{4})$ asociado a
$(P,\pi,M_{4},G)$, y los campos de gauge (que describen la interacción,
como el campo electromagnético), que vienen dados por secciones del
fibrado de conexiones de $(P,\pi,M_{4},G)$; el grupo de Lie $G$
se denomina \emph{grupo de gauge}. En otras palabras: desde este punto de vista ``geométrico'', los campos gauge
mediadores de la interacción son precisamente las conexiones del fibrado principal
 $(P,\pi,M_{4},G)$. De hecho, la ley de
transformación de los representantes locales de las conexiones
bajo un cambio de la sección local que los determina
coincide con la expresión para la
transformacion de los campos de Yang-Mills, y la elección
de un gauge para \'estos corresponde a elegir una seccion
local del fibrado (esta idea fue introducida por C.N. Yang y T. T. Wu en \cite{Yan-Wu 75}. Una referencia
muy detallada, con cálculos explícitos y numerosas aplicaciones, en particular a la relación con las teorías de 
cohomología, es \cite{Azc-Izq 98}). 
Sin embargo, esta estructura introduce
una asimetr\'{\i}a manifiesta entre los campos de materia y los campos
gauge: mientras los primeros pueden ser de tipo bosónico o fermiónico
y están definidos en un fibrado vectorial arbitrario (salvo por estar
asociado a un fibrado principal igualmente arbitrario), los segundos
están definidos en un fibrado de conexiones, y sólo pueden ser de
tipo bosónico. Los términos bosónico y fermiónico se refieren al tipo
de paréntesis de Poisson que es posible definir para estos campos
(cuando se hace un análisis de Fourier en modos cuánticos resultan
los conmutadores o anticonmutadores que ya conocemos para los operadores
de creación-aniquilación), y éste a su vez determina el proceso de
cuantización de los mismos, de modo que las teor\'{\i}as gauge en
su formulación original no pueden verse como las candidatas finales
a una teor\'{\i}a unificada de campos.

Como ya hemos mencionado, las teor\'{\i}as supersimétricas ofrecen
una solución a este problema a través de una sustitución de los conceptos
tradicionales de la Geometr\'{\i}a Diferencial por unos análogos $\mathbb{Z}_{2}-$graduados;
los supercampos que aparecen en esta formulación poseen, en general,
tanto parte una par (bosónica) como una impar (fermiónica), de modo
que las ecuaciones de campo describen simultáneamente campos de materia
y campos de gauge. El principal problema es la complejidad técnica
de su formulación, que hace que no sea sencillo el construir un formalismo
que permita obtener estas ecuaciones de campo de un modo semejante
al clásico; en cualquier caso, el estudio de las variedades $\mathbb{Z}_{2}-$graduadas
o supervariedades, presenta un indiscutible interés no sólo matemático,
sino también f\'{\i}sico (v\'eanse \cite{Fre 86,Wit 92}).

En las siguientes Secciones, describiremos un modelo sencillo que
implementa las ideas básicas de la supersimetr\'{\i}a en el contexto
de la Mecánica Cuántica, siguiendo las ideas de E. Witten (véase \cite{Wit 81}),
que son una reelaboración de otras ya existentes (cfr. \cite{Ber-Mar 75},
\cite{BDZVH 76}). Como referencias b\'asicas citaremos \cite{CKS 01,Cro-Rit 83,Gen-Kri 85,Kib-Dao 04,Fer 09}.

\section{Ejemplo: La superálgebra de Lie $End(\mathbb{C}^{2})$\label{C11}}

Aprovecharemos esta sección para introducir algunas definiciones formales
y analizar un ejemplo muy sencillo pero que reúne todas las características
en las que estamos interesados. Referencias \'utiles para el estudio de las super\'algebras de
Lie son \cite{CNS 75,Sche 79,FSS 00}. 

Consideremos el grupo abeliano $\mathbb{Z}_{2}=\{0,1\}$, con la suma
módulo $2$, y $V$ un $\mathbb{K}-$espacio vectorial. Se dice que
$V$ es $\mathbb{Z}_{2}-$graduado, ó que es un $\mathbb{K}-$superespacio
vectorial, si admite una descomposición en suma directa de subespacios\[
V=\bigoplus_{m\in\mathbb{Z}_{2}}V_{m}=V_{0}\oplus V_{1}.\]
Los elementos del subespacio $V_{m}$ se llaman homogéneos de grado
$m\in\mathbb{Z}_{2}$. Cuando $m=0$ se dice que son elementos \emph{pares}
y cuando $m=1$, \emph{impares}. Se define así una aplicación 
$|\cdot|:(V_1 \cup V_2 )-\{0\}\rightarrow\mathbb{Z}_{2}$
mediante\[
|v|=m\; si\; v\in V_{m},\]
llamada aplicación \emph{grado}. Con el fin de que esta aplicación
resulte bien definida, se conviene en asignar a los elementos de $\mathbb{K}$
grado $0$ y establecer que $|k.v|$ (con $k\in\mathbb{K}$ y $v\in V$)
es $|k|+|v|=|v|$.

Supongamos ahora que $V$ y $W$ son $\mathbb{K}-$superespacios vectoriales.
Una aplicación $\mathbb{K}-$lineal $\phi:V\rightarrow W$ se dice
que es homogénea de grado $p\in\mathbb{Z}_{2}$ (par si $p=0$, impar
si $p=1$) cuando ocurre que\[
\phi(V_{m})\subset W_{m+p}\]
para todo $m\in\mathbb{Z}_{2}$. Si $p=0$ (esto es, $\phi$ es una
aplicación par) suele decirse que $\phi\in Hom_{\mathbb{K}}(V,W)$
es un morfismo de $\mathbb{K}-$superespacios vectoriales. Naturalmente,
se tiene el caso particular en que $V=W$. Entonces, se habla de endomorfismos
homogéneos de grado $p\in\mathbb{Z}_{2}$ y de endomorfismos de superespacios
vectoriales si $p=0$.

Recordemos que una $\mathbb{K}-$álgebra es un $\mathbb{K}-$espacio
vectorial $A$ junto con una aplicación $\mathbb{K}-$bilineal $A\times A\rightarrow A$
que escribiremos como $(a_{1},a_{2})\mapsto a_{1}\cdot a_{2}$. Si
$A$ es un $\mathbb{K}-$superespacio vectorial y esta aplicación
lleva $A_{m}\times A_{n}$ en $A_{m+n}$ (la suma módulo $2$), con
$m,n\in\mathbb{Z}_{2}$, entonces diremos que $A$ es una $\mathbb{K}-$álgebra
$\mathbb{Z}_{2}-$graduada o, simplemente, una superálgebra.

Como veremos enseguida, puede darse el caso en que una superálgebra
$A$ tenga, además, definida otra operación, una aplicación $\mathbb{K}-$bilineal
$[\![\cdot,\cdot]\!]:A\times A\rightarrow A$ tal que se cumplen las
siguientes propiedades adicionales:

\begin{enumerate}
\item $[\![a_{1},a_{2}]\!]=-(-1)^{|a_{1}||a_{2}|}[\![a_{2},a_{1}]\!]$ (antisimetría
$\mathbb{Z}_{2}-$graduada),
\item $[\![a_{1},[\![a_{2},a_{3}]\!]]\!]=[\![[\![a_{1},a_{2}]\!],a_{3}]\!]+(-1)^{|a_{1}||a_{2}|}[\![a_{2},[\![a_{1},a_{3}]\!]]\!]$
(identidad de Jacobi $\mathbb{Z}_{2}-$graduada).
\end{enumerate}
En tal caso, se dice que $(A,[\![\cdot,\cdot]\!])$ forma una superálgebra
de Lie. La operación $[\![\cdot,\cdot]\!]$ suele denominarse corchete
de Lie (graduado) o supercorchete.

Veamos cómo surgen de manera natural las superálgebras de Lie. Consideremos
un superespacio vectorial $V=V_{0}\oplus V_{1}$ y los endomorfismos
homogéneos sobre $V$, $End_{\mathbb{K}}^{G}(V)$. Sabemos que se
tiene la descomposición inducida por el grado\[
End_{\mathbb{K}}^{G}(V)=End_{\mathbb{K}}^{0}(V)\oplus End_{\mathbb{K}}^{1}(V),\]
donde\[
\begin{array}{c}
End_{\mathbb{K}}^{0}(V)=\{f\in End_{\mathbb{K}}^{G}(V):f(V_{i})\subset V_{i}\}\\
\\End_{\mathbb{K}}^{1}(V)=\{f\in End_{\mathbb{K}}^{G}(V):f(V_{i})\subset V_{(i+1)mod2}\}\end{array}\]
siendo $i\in\{0,1\}$. En este espacio de endomorfismos homogéneos,
está definida la composición%
\footnote{De manera que $(End_{\mathbb{K}}^{G}(V),\circ)$ es una superálgebra
si se establece que $|f+g|=|f|+|g|$. Se deja la comprobación como
un sencillo ejercicio.%
} y, a partir de ella, construímos un corchete de Lie graduado $[\![\cdot,\cdot]\!]:End_{\mathbb{K}}^{G}(V)\times End_{\mathbb{K}}^{G}(V)\rightarrow End_{\mathbb{K}}^{G}(V)$
poniendo\[
[\![f,g]\!]=f\circ g-(-1)^{|f||g|}g\circ f.\]
Es inmediato%
\footnote{Ejercicio para el lector.%
} comprobar que esta definición determina una estructura de superálgebra
de Lie sobre $End_{\mathbb{K}}^{G}(V)$. Este ejemplo de superálgebra
de Lie es universal en el siguiente sentido: al igual que ocurre en
la teoría de álgebras de Lie clásicas, donde todo álgebra $n-$dimensional
sobre un cuerpo $\mathbb{K}$, $\mathfrak{g}$, es isomorfa a una
del tipo $(End_{\mathbb{K}}(\mathbb{K}^{n}),[\cdot,\cdot])$ con $[\cdot,\cdot]$
el conmutador de endomorfismos dado por $[A,B]=A\circ B-B\circ A$
(teorema de Ado), en el caso $\mathbb{Z}_{2}-$graduado toda superálgebra
también es isomorfa a una del tipo $(End_{\mathbb{K}}^{G}(V),[\![\cdot,\cdot]\!])$
(véase \cite{Sche 79} $\S$4). De hecho, el objetivo
central de este trabajo consiste en mostrar explícitamente cómo el
álgebra de la Mecánica Cuántica supersimétrica se representa en esta
forma. Obviamente, para lograr este objetivo necesitaremos primero
dar el superespacio vectorial $V=V_{0}\oplus V_{1}$ sobre el cual
se considerarán los endomorfismos graduados homogéneos.

Como un ejercicio previo, vamos a estudiar la estructura de $End_{\mathbb{C}}^{G}(V)$
cuando se tiene el $\mathbb{C}-$superespacio vectorial $\mathbb{C}^{1|1}=\mathbb{C}\oplus\Pi\mathbb{C}$,
donde $\mathbb{C}\simeq\Pi\mathbb{C}$ son isomorfos y la única diferencia
entre ellos es que la copia $\mathbb{C}$ corresponde a los elementos
de $\mathbb{C}^{1|1}$ con grado $0$ (i.e, $\mathbb{C}=(\mathbb{C}^{1|1})_{0}$)
y la copia $\Pi\mathbb{C}$ contienen a los elementos de $\mathbb{C}^{1|1}$
con grado $1$ (i.e, $\Pi\mathbb{C}=(\mathbb{C}^{1|1})_{1}$). En
otras palabras: todo elemento $z\in\mathbb{C}^{1|1}$ se escribe en
la forma $z=z_{0}\oplus z_{1}$, con $z_{0}\in\mathbb{C}$ y $z_{1}\in\Pi\mathbb{C}$,
siendo pues $|z_{0}|=0$ y $|z_{1}|=1$. Sin embargo, ya que la manera
más cómoda de trabajar con endomorfismos consiste en utilizar sus
representantes matriciales, aprovecharemos que como $\mathbb{C}-$espacios
vectoriales $\mathbb{C}^{1|1}\simeq\mathbb{C}^{2}$ e introduciremos
la notación siguiente: si $z\in\mathbb{C}^{1|1}=\mathbb{C}\oplus\Pi\mathbb{C}$
se descompone como $z=z_{0}\oplus z_{1}$ escribiremos\[
z=\left(\begin{array}{c}
z_{0}\\
z_{1}\end{array}\right)\]
(esto es, vemos a los elementos de $\mathbb{C}^{1|1}$ como biespinores).
Con esta notación, un $f\in End_{\mathbb{C}}^{G}(\mathbb{C}^{1|1})$
se escribirá como\[
f=\left(\begin{array}{cc}
f_{00} & f_{01}\\
f_{10} & f_{11}\end{array}\right),\]
y su acción vendrá dada por\[
f(z)=\left(\begin{array}{cc}
f_{00} & f_{01}\\
f_{10} & f_{11}\end{array}\right)\left(\begin{array}{c}
z_{0}\\
z_{1}\end{array}\right)=\left(\begin{array}{c}
f_{00}z_{0}+f_{01}z_{1}\\
f_{10}z_{0}+f_{11}z_{1}\end{array}\right).\]
Ahora bien, dado que $f_{00}z_{0}+f_{01}z_{1}$ debe tener grado $0$
y $f_{10}z_{0}+f_{11}z_{1}$ grado $1$, mientras que $|z_{0}|=0$
y $|z_{1}|=1$, para que $f$ sea homogénea de grado $0$ se requiere
que\[
\begin{array}{c}
|f_{00}|=0=|f_{11}|\\
|f_{10}|=1=|f_{01}|\end{array}.\]
Pero los elementos de matriz del endomorfismo $f\in End_{\mathbb{C}}^{G}(\mathbb{C}^{1|1})$
son elementos del cuerpo base $\mathbb{C}$ y estos elementos recordemos
que siempre tienen grado $0$. En consecuencia, podemos escribir\[
End_{\mathbb{C}}^{0}(\mathbb{C}^{1|1})=\left\{ f\in End_{\mathbb{C}}^{G}(\mathbb{C}^{1|1}):f\equiv\left(\begin{array}{cc}
f_{00} & 0\\
0 & f_{11}\end{array}\right)\right\} .\]

Un análisis análogo nos conduce a\[
End_{\mathbb{C}}^{1}(\mathbb{C}^{1|1})=\left\{ f\in End_{\mathbb{C}}^{G}(\mathbb{C}^{1|1}):f\equiv\left(\begin{array}{cc}
0 & f_{01}\\
f_{10} & 0\end{array}\right)\right\} .\]

Es útil, por otra parte, observar lo siguiente: dada la base canónica
de $\mathbb{C}^{1|1}$,\[
\mathcal{B}=\left\{ e_{0}=\left(\begin{array}{c}
1\\
0\end{array}\right),e_{1}=\left(\begin{array}{c}
0\\
1\end{array}\right)\right\} ,\]
automáticamente se tiene una base asociada en $End_{\mathbb{C}}^{G}(\mathbb{C}^{1|1})$,
dada por%
\footnote{En este contexto, $\Phi,\Psi^{\dagger},\Psi,\Phi^{\dagger}$ son simplemente
unos nombres para los vectores base de $End_{\mathbb{C}}^{G}(\mathbb{C}^{1|1})$.
En particular, $\dagger$ no denota conjugación de ningún tipo (y,
de hecho, más adelante veremos otra convención igualmente extendida
para denominar a estos elementos utilizando los símbolos $\sigma_{0},...,\sigma_{3}$).
El origen de esta notación está en su similitud con otros operadores
en Física que si están relacionados por cierto tipo de conjugación.%
} \[
\mathcal{\widetilde{B}}=\left\{ \Phi=\left(\begin{array}{cc}
1 & 0\\
0 & 0\end{array}\right),\Psi^{\dagger}=\left(\begin{array}{cc}
0 & 1\\
0 & 0\end{array}\right),\Psi=\left(\begin{array}{cc}
0 & 0\\
1 & 0\end{array}\right),\Phi^{\dagger}=\left(\begin{array}{cc}
0 & 0\\
0 & 1\end{array}\right)\right\} .\]
De acuerdo con lo dicho, de entre los elementos de la base $\mathcal{\widetilde{B}}$
hay dos pares:\[
|\Phi|=\left|\left(\begin{array}{cc}
1 & 0\\
0 & 0\end{array}\right)\right|=0=\left|\left(\begin{array}{cc}
0 & 0\\
0 & 1\end{array}\right)\right|=|\Phi^{\dagger}|,\]
y dos impares:\[
|\Psi|=\left|\left(\begin{array}{cc}
0 & 0\\
1 & 0\end{array}\right)\right|=1=\left|\left(\begin{array}{cc}
0 & 1\\
0 & 0\end{array}\right)\right|=|\Psi^{\dagger}|.\]

Todo elemento $f\in End_{\mathbb{C}}^{G}(\mathbb{C}^{1|1})$ se escribe
como una combinación $\mathbb{C}-$lineal\[
f=f_{00}\Phi+f_{01}\Psi^{\dagger}+f_{10}\Psi+f_{11}\Phi^{\dagger}.\]
Si $f\in End_{\mathbb{C}}^{0}(\mathbb{C}^{1|1})$, entonces su expresión
coordenada se reduce a\[
f=f_{00}\Phi+f_{11}\Phi^{\dagger},\]
y si es $g\in End_{\mathbb{C}}^{1}(\mathbb{C}^{1|1})$, entonces,\[
g=g_{01}\Psi^{\dagger}+g_{10}\Psi.\]

Como es bien sabido, la composición de endomorfismos se corresponde
con el producto matricial de sus representantes matriciales. Esto
hace que los cálculos con la estructura de superálgebra de Lie $(End_{\mathbb{C}}^{1}(\mathbb{C}^{1|1}),[\![\cdot,\cdot]\!])$
sean muy sencillos. Por ejemplo, supongamos que\[
f=\left(\begin{array}{cc}
e^{i\alpha} & 0\\
0 & 0\end{array}\right),g=\left(\begin{array}{cc}
0 & e^{i\beta}\\
3e^{i\gamma} & 0\end{array}\right),\alpha,\beta,\gamma\in\mathbb{R}.\]
Entonces:
\begin{eqnarray*}
[\![f,g]\!]&=& 
\left(\begin{array}{cc}
e^{i\alpha} & 0\\
0 & 0\end{array}\right)\left(\begin{array}{cc}
0 & e^{i\beta}\\
3e^{i\gamma} & 0\end{array}\right)-(-1)^{0\cdot1}\left(\begin{array}{cc}
0 & e^{i\beta}\\
3e^{i\gamma} & 0\end{array}\right)\left(\begin{array}{cc}
e^{i\alpha} & 0\\
0 & 0\end{array}\right) \\
&=& \left(\begin{array}{cc}
0 & e^{i(\alpha+\beta)}\\
-3e^{i(\alpha+\gamma)} & 0
\end{array}\right)
\end{eqnarray*}
y claramente $|[\![f,g]\!]|=1$ (no hace falta fijarse en la forma
de la matriz resultante, basta con concocer los grados de los endomorfismos
$f$ y $g$: observemos que el grado del corchete siempre es el grado
$|f\circ g|=|f|+|g|$).

\section{Mecánica Cuántica Supersimétrica (SUSY QM)}

Hemos mencionado en repetidas ocasiones que uno de nuestros objetivos es el de dar
una realización de la superálgebra de Heisenberg. En esta sección vamos a ver cuál
es este álgebra y analizaremos su origen físico tal y como suele presentarse en
los textos de Física, basándonos en el formalismo de los operadores de creación 
y aniquilación. Una vez que sepamos cómo es el álgebra y su significado físico, 
resultará más sencillo formalizar su construcción en el contexto de los endomorfismos
de un superespacio vectorial (esto es, en el contexto de las supermatrices).

Supongamos una part\'{\i}cula que posee un grado de libertad bosónico
y un grado de libertad fermiónico, esto es, para caracterizar el estado
de la part\'{\i}cula supondremos que debemos dar dos vectores correspondientes
respectivamente a los espacios de Hilbert de variables ``bosónicas''
y ``fermiónicas''. En otras palabras, el espacio de Hilbert que
describe al sistema será
\[
\mathcal{H}=\mathcal{H}_{B}\otimes\mathcal{H}_{F}\text{.}
\]
 Por ejemplo, para una part\'{\i}cula de espín $\frac{1}{2}$ moviéndose
en una dimensión la posición de la part\'{\i}cula es el grado de libertad
bosónico, y $\mathcal{H}_{B}=L^{2}(\mathbb{R})$, mientras que según
hemos visto en la sección \ref{spin1/2}, $\mathcal{H}_{F}=\mathbb{C}^{2}$.
En el caso de tener $N$ part\'{\i}culas (de las cuales hay $n_{B}$
bosones y $n_{F}$ fermiones) de acuerdo con el Principio de Pauli
debemos simetrizar los estados bosónicos y antisimetrizar los fermiónicos,
es decir, debemos tomar
\[
\mathcal{H}=S\mathcal{H}_{B}\otimes\Lambda\mathcal{H}_{F},
\]
 y podemos trabajar entonces con la representación por el número de
ocupación. As\'{\i}, para describir el sistema necesitamos dar un
vector 
\[
\begin{array}{c}
\left\vert n_{B}\right\rangle \equiv\left\vert \mu_{1}\mu_{2}...\mu_{r}\right\rangle \in S\mathcal{H}_{B}\text{ con }\mu_{i}\in\mathbb{N},1\leq i\leq r,\mu_{1}+\cdots+\mu_{r}=n_{B}\\
\left\vert n_{F}\right\rangle \equiv\left\vert \nu_{1}\nu_{2}...\nu_{s}\right\rangle \in\Lambda\mathcal{H}_{B}\text{ con }\nu_{j}\in\{0,1\},1\leq j\leq s,\mbox{ }\nu_{1}+\cdots+\nu_{s}=n_{F}.
\end{array}
\]

Para un elemento arbitrario, pondremos
\[
\left\vert n_{B}\right\rangle \otimes\left\vert n_{F}\right\rangle \equiv\left\vert n_{B},n_{F}\right\rangle .
\]
 En este contexto, se pueden introducir heur\'{\i}sticamente unos
operadores de creación y aniquilación para estados bosónicos y fermiónicos
de la siguiente forma. Para bosones tenemos (obsérvese la notación
abreviada): 
\begin{equation}
\left.
\begin{array}{r}
a_{r}\left\vert n_{B},n_{F}\right\rangle =\left(a_{r}\otimes\mathbf{1}\right)\left\vert \mu_{1}...\mu_{r}...\right\rangle \otimes\left\vert \nu_{1}...\nu_{s}...\right\rangle \\
=\left\vert \mu_{1}...\mu_{r}-1...\right\rangle \otimes\left\vert \nu_{1}...\nu_{s}...\right\rangle =\left\vert n_{B}-1_{r},n_{F}\right\rangle \\
\multicolumn{1}{c}{\text{y}}\\
a_{r}^{\dag}\left\vert n_{B},n_{F}\right\rangle =\left(a_{r}^{\dag}\otimes\mathbf{1}\right)\left\vert \mu_{1}...\mu_{r}...\right\rangle \otimes\left\vert \nu_{1}...\nu_{s}...\right\rangle \\
=\left\vert \mu_{1}...\mu_{r}+1...\right\rangle \otimes\left\vert \nu_{1}...\nu_{s}...\right\rangle =\left\vert n_{B}+1_{r},n_{F}\right\rangle ,
\end{array}\right\} \label{eq0_11}
\end{equation}
 donde $n_{B}=\mu_{1}+...+\mu_{r}+...,$ $n_{F}=\nu_{1}+...+\nu_{s}+...$,
y para fermiones: 
\begin{equation}
\left.
\begin{array}{c}
b_{s}\left\vert n_{B},n_{F}\right\rangle =\left(\mathbf{1}\otimes b_{s}\right)\left\vert \mu_{1}...\mu_{r}...\right\rangle \otimes\left\vert \nu_{1}...\nu_{s}...\right\rangle \\
=\left\vert \mu_{1}...\mu_{r}...\right\rangle \otimes\left\vert \nu_{1}...\nu_{s}-1...\right\rangle =\left\vert n_{B},n_{F}-1_{s}\right\rangle \\
\text{y}\\
b_{s}^{\dag}\left\vert n_{B},n_{F}\right\rangle =\left(\mathbf{1}\otimes b_{s}^{\dag}\right)\left\vert \mu_{1}...\mu_{r}...\right\rangle \otimes\left\vert \nu_{1}...\nu_{s}...\right\rangle \\
=\left\vert \mu_{1}...\mu_{r}...\right\rangle \otimes\left\vert \nu_{1}...\nu_{s}+1...\right\rangle =\left\vert n_{B},n_{F}+1_{s}\right\rangle .
\end{array}
\right\} \label{eq0_12}
\end{equation}
 Naturalmente, es inmediato que estos operadores cumplen las siguientes
relaciones algebraicas:
\begin{align*}
[a_{j},a_{k}]\_ = \delta_{jk} \\
[b_{r},b_{s}]_{+} = \delta_{rs} \\
[a_{j},b_{s}]\_ = 0,
\end{align*}
(donde realmente habr\'{\i}a que escribir $[a_{j}\otimes\mathbf{1},a_{k}\otimes\mathbf{1}]\_ =\delta_{jk}\mathbf{1}\otimes\mathbf{1}$,
etc, pero el contexto deber\'{\i}a evitar las confusiones).

Para introducir la supersimetr\'{\i}a, querr\'{\i}amos disponer de
operadores que transformasen un estado bosónico en uno fermiónico
y viceversa. De las expresiones (\ref{eq0_11}) y (\ref{eq0_12})
se ve que basta con definir unos operadores adjunto el uno del otro
como sigue:
\[
\begin{array}{c}
Q_{rs}=a_{r}b_{s}^{\dag}=a_{r}\otimes b_{s}^{\dag}\\
Q_{rs}^{\dag}=a_{r}^{\dag}b_{s}=a_{r}^{\dag}\otimes b_{s}.
\end{array}
\]
 En efecto:
 \begin{equation}
\begin{array}{r}
Q_{rs}\left\vert n_{B},n_{F}\right\rangle =a_{r}\left\vert \mu_{1}...\mu_{r}...\right\rangle \otimes b_{s}^{\dag}\left\vert \nu_{1}...\nu_{s}...\right\rangle \\
=\left\vert \mu_{1}...\mu_{r}-1...\right\rangle \otimes\left\vert \nu_{1}...\nu_{s}+1...\right\rangle \\
=\left\vert n_{B}-1_{r},n_{F}+1\right\rangle ,
\end{array}\label{eq0_13}
\end{equation}
 y de manera similar 
\begin{equation}
Q_{rs}^{\dag}\left\vert n_{B},n_{F}\right\rangle =\left\vert n_{B}+1_{r},n_{F}-1\right\rangle .\label{eq0_14}
\end{equation}

Se tiene una relación important\'{\i}sima entre los operadores $Q_{rs}$
y $Q_{rs}^{\dag}$. Si se calcula su anticonmutador, aplicando las
reglas de conmutación para las $a$'s y las $b$'s, resulta: 
\begin{eqnarray*}
[Q_{rs}^{\dag},Q_{rs}]_{+} & = & [a_{r}^{\dag}\otimes b_{s},a_{r}\otimes b_{s}^{\dag}]_{+}\\
 & = & \left(a_{r}^{\dag}\otimes b_{s}\right)\left(a_{r}\otimes b_{s}^{\dag}\right)+\left(a_{r}\otimes b_{s}^{\dag}\right)\left(a_{r}^{\dag}\otimes b_{s}\right)\\
 & = & a_{r}^{\dag}a_{r}\otimes b_{s}b_{s}^{\dag}+a_{r}a_{r}^{\dag}\otimes b_{s}^{\dag}b_{s}\\
 & = & a_{r}^{\dag}a_{r}\otimes\left(\mathbf{1}-b_{s}b_{s}^{\dag}\right)+\left(\mathbf{1}+a_{r}a_{r}^{\dag}\right)\otimes b_{s}^{\dag}b_{s}\\
 & = & a_{r}^{\dag}a_{r}\otimes\mathbf{1}-a_{r}^{\dag}a_{r}\otimes b_{s}b_{s}^{\dag}+a_{r}a_{r}^{\dag}\otimes b_{s}^{\dag}b_{s}+\mathbf{1}\otimes b_{s}^{\dag}b_{s}\\
 & = & n_{r}^{B}\otimes\mathbf{1}+\mathbf{1}\otimes n_{s}^{F},
 \end{eqnarray*}
 es decir: $[Q_{rs}^{\dag},Q_{rs}]_{+}$ nos da el número de bosones
en el estado $n_{r}$ y el de fermiones en el estado $n_{s}$. As\'{\i}
\begin{equation}
\sum\limits _{r,s}[Q_{rs}^{\dag},Q_{rs}]_{+}=N_{B}\otimes\mathbf{1}+\mathbf{1}\otimes N_{F},\label{eq0_14b}
\end{equation}
 donde $N_{B}$ es el \emph{operador número total de bosones} y $N_{F}$
es el \emph{operador número total de fermiones}: 
\begin{eqnarray*}
N_{B}\otimes\mathbf{1}\left\vert n_{B},n_{F}\right\rangle  & = & \left(\mu_{1}+...+\mu_{r}+...\right)\left\vert n_{B},n_{F}\right\rangle =n_{B}\left\vert n_{B},n_{F}\right\rangle \\
\mathbf{1}\otimes N_{F}\left\vert n_{B},n_{F}\right\rangle  & = & \left(\nu_{1}+...+\nu_{s}+...\right)\left\vert n_{B},n_{F}\right\rangle =n_{F}\left\vert n_{B},n_{F}\right\rangle .
\end{eqnarray*}
Por tanto,\[
\left(N_{B}\otimes\mathbf{1}+\mathbf{1}\otimes N_{F}\right)\left\vert n_{B},n_{F}\right\rangle =(n_{B}+n_{F})\left\vert n_{B},n_{F}\right\rangle =N\left\vert n_{B},n_{F}\right\rangle \]
y debido a esto, al operador $N=N_{B}\otimes\mathbf{1}+\mathbf{1}\otimes N_{F}$
se le denomina \emph{operador número total de partículas}.

Por otra parte, el hecho de que tanto $Q_{rs}^{\dag}$ como $Q_{rs}$
contengan a los operadores fermiómicos $b_{s}^{\dag},b_{s}$, hace
que sean nilpotentes: 
\begin{equation}
\begin{array}{c}
\left(Q_{rs}^{\dag}\right)^{2}\left\vert n_{B},n_{F}\right\rangle =\left(a_{r}^{\dag}\right)^{2}\left\vert n_{B}\right\rangle \otimes\left(b_{s}\right)^{2}\left\vert n_{F}\right\rangle =\left(a_{r}^{\dag}\right)^{2}\left\vert n_{B}\right\rangle \otimes\mathbf{0}=0\\
\left(Q_{rs}\right)^{2}\left\vert n_{B},n_{F}\right\rangle =\left(a_{r}\right)^{2}\left\vert n_{B}\right\rangle \otimes\left(b_{s}^{\dag}\right)^{2}\left\vert n_{F}\right\rangle =\left(a_{r}\right)^{2}\left\vert n_{B}\right\rangle \otimes\mathbf{0}=0.
\end{array}\label{eq0_15}
\end{equation}
 Esta propiedad de nilpotencia es la que nos permitirá construir un
Hamiltoniano que admita a (\ref{eq0_13}) y (\ref{eq0_14}) como transformaciones
de simetr\'{\i}a%
\footnote{Conviene recordar que, en Mecánica Cuántica, un operador $A$ se dice
que es el generador de unas transformaciones de simetr\'{\i}a si conmuta
con el Hamiltoniano del sistema. Esto está relacionado con el hecho
de que la ecuación de evolución para $A$ sea $\frac{dA}{dt}=[A,H]$
y, por tanto, $[A,H]=0$ implica $A$ constante del movimiento (si $A$ no depende
explícitamente del tiempo).} y que, por mezclar bosones y fermiones, denominaremos
\emph{supersimetr\'{\i}as}.

Basta con tomar 
\begin{equation}
H=\sum\limits _{r,s}[Q_{rs}^{\dag},Q_{rs}]_{+}=N_{B}\otimes\mathbf{1}+\mathbf{1}\otimes N_{F}\label{eq0_16}
\end{equation}
 y fijarse en que para cualquier par de \'{\i}ndices $j,k$: 
 \begin{equation}
[H,Q_{jk}^{\dag}]\_ =0=[H,Q_{jk}]\_ .\label{eq0_17}
\end{equation}
 Por ejemplo: 
 \begin{eqnarray*}
[H,Q_{jk}]\_ & = & \sum\limits _{r,s}[[Q_{rs}^{\dag},Q_{rs}]_{+} ,Q_{jk}]\_ \\
 & = & \sum\limits _{r,s}[[a_{r}^{\dag}b_{s},a_{r}b_{s}^{\dag}]_{+},a_{j}b_{k}^{\dag}]\_ \\
 & = & \sum\limits _{r,s}[[a_{r}^{\dag},a_{r}]_{+},a_{j}]\_ \otimes[[b_{s},b_{s}^{\dag}]_{+},b_{k}^{\dag}]\_ \\
 & = & \sum\limits _{r,s}[[a_{r}^{\dag},a_{r}]_{+},a_{j}]\_ \otimes[\mathbf{1},b_{k}^{\dag}]\_ \\
 & = & \sum\limits _{r,s}[[a_{r}^{\dag},a_{r}]_{+},a_{j}]\_ \otimes0=0.
 \end{eqnarray*}

Observemos ahora otra propiedad básica%
\footnote{Ejercicio. No hay más que repetir el razonamiento de (\ref{eq0_14b}).%
} del anticonmutador de los operadores $Q_{rs}^{\dag}\ $y $Q_{rs}$:
\begin{equation}
[Q_{jk}^{\dag},Q_{rs}]_{+}=0\label{eq0_18}
\end{equation}
 cuando $\left(r,s\right)\neq\left(j,k\right)$. A la vista de (\ref{eq0_15}),
(\ref{eq0_16}), (\ref{eq0_17}) y (\ref{eq0_18}), definiendo
\begin{equation}
\begin{array}{c}
Q=\sum\limits _{r,s}Q_{rs}\\
Q^{\dag}=\sum\limits _{r,s}Q_{rs}^{\dag},
\end{array}\label{eq0_18b}
\end{equation}
 resulta claro que el espacio vectorial sobre $\mathbb{C}$ generado
por $\{H,Q,Q^{\dag}\}$ tiene la estructura de una superálgebra de
Lie:
\begin{align}
[Q^{\dagger},Q]_{+} &= \sum\limits _{r,s}[Q_{rs}^{\dag},Q_{rs}]_{+}=H\notag\\
[H,H]\_ &= 0\label{eq0_19}\\
[Q,H]\_ &= 0\notag\\
[Q^{\dag},H]\_ &= 0\notag
\end{align}
(en la primera ecuación se ha utilizado (\ref{eq0_18}) para eliminar
la contribución de los términos cruzados). Para verlo aún más expl\'{\i}citamente,
declaramos que el grado de cada operador viene dado por
\begin{eqnarray*}
\left\vert H\right\vert  & = & 0\\
\left\vert Q\right\vert  & = & 1=\left\vert Q^{\dag}\right\vert ,
\end{eqnarray*}
 y agrupamos los conmutadores $[.,.]\_$,$[.,.]_{+}$ en uno solo $[\![.,.]\!]$,
de manera que $[\![.,.]\!]$ sobre dos elementos impares actúa como
$[.,.]_{+}$ y coincide con $[.,.]\_$ cuando hay un elemento par. Si
$F,G\in\{H,Q,Q^{\dag}\}$ podemos poner, entonces,\[
[\![F,G]\!]=F\circ G-(-1)^{FG}G\circ F,\]
 como en (\ref{eq0_10}), y las ecuaciones (\ref{eq0_19}) se resumen entonces en
 la llamada \emph{superálgebra de Heisenberg}:
\begin{gather}
[\![Q^{\dag},Q]\!]=H \nonumber \\
[\![Q^{\dag},H]\!]=0=[\![Q,H]\!].\label{salgebra}
\end{gather}

Fijémonos, por otra parte, en que la ecuación (\ref{eq0_16}) puede
interpretarse f\'{\i}sicamente como la suma del Hamiltoniano de un
oscilador armónico bosónico y otro oscilador fermiónico, sin interacción
entre ellos. Lo que es importante en este punto, es que la existencia
de supersimetr\'{\i}as se mantiene incluso cuando se introduce una
interacción. Ciertamente, podr\'{\i}amos generalizar las definiciones
de $Q_{rs}^{\dag},Q_{rs}$ y poner
\begin{eqnarray*}
Q_{rs}^{\dag} & = & B_{r}^{\dag}\left(a_{r}^{\dag},a_{r}\right)b_{s}\\
Q_{rs} & = & B_{r}\left(a_{r}^{\dag},a_{r}\right)b_{s}^{\dag},
\end{eqnarray*}
 donde $B,B^{\dag}$ son funciones arbitrarias%
\footnote{Observemos que si $Q_{rs}^{\dag}$ es el adjunto de $Q_{rs}$, esto
implica que $B_{r}^{\dag}$ debe ser el adjunto de $B_{r}$. La arbitrariedad
se entiende salvo este hecho.%
} de los operadores bosónicos $a_{r}^{\dag},a_{r}$. Definiendo entonces
$Q,Q^{\dag}$ y $H$ como antes (véase (\ref{eq0_18b})), las propiedades
(\ref{eq0_19}) se siguen cumpliendo, y se sigue teniendo una estructura
supersimétrica. Sin embargo, no seguiremos este camino. En la próxima sección 
veremos otra descripción, más cercana al formalismo de las superálgebras de Lie,
en la que partiremos directamente de la ecuación de Schrödinger incluyendo el
potencial de interacción para llegar a la superálgebra (\ref{salgebra}). Como se
verá en la última sección, esta nueva descripción resulta mucho más útil por lo 
que respecta a las aplicaciones.

\section{Realización explícita de la superálgebra de Heisenberg}

En esta sección, vamos a realizar la superálgebra $\{H,Q,Q^{\dag}\}$
como subálgebra de los endomorfismos de un cierto superespacio
vectorial. Este espacio no es otro que el producto tensorial de las
funciones de cuadrado integrable con el la suma directa de dos copias
del plano complejo, donde a la primera copia se le asigna paridad
$0$ y a la segunda paridad $1$.

Consideraremos un sistema con bosones y fermiones de spin $\frac{1}{2}$
y supondremos que la evolución del sistema está determinada, de acuerdo
con lo expuesto en la sección \ref{S5}, por una ecuación
de Schrödinger que incluye un potencial de interacción entre las
partículas $V$. El espacio de las funciones de onda%
\footnote{Los bosones también se describen mediante funciones de onda de $\mathcal{H}$
pero sólo con aquellas pertenecientes al subespacio $L^{2}(\mathbb{R})\otimes\left\langle \left(\begin{array}{c}
1\\
0\end{array}\right)\right\rangle $.%
} será $\mathcal{H}=L^{2}(\mathbb{R})\otimes\mathbb{C}^{1|1}$.

Ahora, $\mathcal{H}$ adquiere una estructura de $\mathbb{C}-$superespacio
vectorial asignándole al factor $L^{2}(\mathbb{R})$ la gradación
(par) trivial, de modo que utilizando la distributividad del producto
tensorial sobre la suma directa de espacios vectoriales se tiene la
descomposición\[
\mathcal{H}=\mathcal{H}_{0}\oplus\mathcal{H}_{1}=(L^{2}(\mathbb{R})\otimes\mathbb{C})\oplus(L^{2}(\mathbb{R})\otimes\Pi\mathbb{C}).\]

Veamos con un poco más de detalle la estructura de $\mathcal{H}=\mathcal{H}_{0}\oplus\mathcal{H}_{1}$.
Dada una función de onda $\psi(x)\in L^{2}(\mathbb{R})$ y un elemento\[
z=z_{0}\oplus z_{1}=\left(\begin{array}{c}
z_{0}\\
z_{1}\end{array}\right)\in\mathbb{C}^{1|1},\]
el producto tensorial $\psi(x)\otimes z$ puede representarse matricialmente
mediante\[
\psi(x)\otimes z=\left(\begin{array}{c}
\psi(x)\cdot z_{0}\\
\psi(x)\cdot z_{1}\end{array}\right)=\left(\begin{array}{c}
\psi(x)\cdot z_{0}\\
0\end{array}\right)\oplus\left(\begin{array}{c}
0\\
\psi(x)\cdot z_{1}\end{array}\right),\]
esto es,\[
\psi(x)\otimes z=(\psi(x)\otimes z_{0})\oplus(\psi(x)\otimes z_{1}).\]

Supongamos ahora que tenemos el Hamiltoniano clásico\[
\widetilde{H}=-\frac{1}{2}\frac{d^{2}}{dx^{2}}+V(x)\]
donde, por simplicidad, omitimos el factor constante $\frac{\hbar^{2}}{m}$
que es inesencial para nuestra discusión. El lector puede, como ejercicio,
restaurar este factor en cada una de las expresiones que siguen (en
términos físicos, estamos utilizando unidades naturales, $\hbar=1$,
y hemos rescalado la unidad de masa para que $m=1$). Nótese que denominamos
este Hamiltoniano con $\widetilde{H}$ y no simplemente $H$, el motivo
es que más adelante construiremos el superhamiltoniano $H$ a partir
de $\widetilde{H}$. Queremos hallar una realización del superálgebra
de Heisenberg $[\![Q^{\dagger},Q]\!]=H$ en términos de operadores
pertenecientes a los endomorfismos homogéneos de algún superespacio
vectorial. El candidato natural es el superespacio $\mathcal{H}=\mathcal{H}_{0}\oplus\mathcal{H}_{1}$
que acabamos de introducir; por otra parte, en la superálgebra de
Heisenberg $H$ debe tener grado $|H|=0$ y los elementos de $End_{\mathbb{C}}^{G}(\mathcal{H})$
tienen la forma $T\otimes f$, donde $T\in End_{\mathbb{C}}(L^{2}(\mathbb{R}))$
y $f\in End_{\mathbb{C}}^{G}(\mathbb{C}^{1|1})$, de modo que se pueden
representar como\begin{equation}
\begin{array}{c}
T\otimes f=\left(\begin{array}{cc}
f_{00}\cdot T & 0\\
0 & f_{11}\cdot T\end{array}\right)\quad si\; f\in End_{\mathbb{C}}^{0}(\mathbb{C}^{1|1})\\
\\T\otimes f=\left(\begin{array}{cc}
0 & f_{01}\cdot T\\
f_{10}\cdot T & 0\end{array}\right)\quad si\; f\in End_{\mathbb{C}}^{1}(\mathbb{C}^{1|1})\end{array}.\label{eq:Tyf}\end{equation}
Como acabamos de decir, $H$ debe pertenecer al primer tipo, mientras
que los operadores $Q$ y $Q^{\dagger}$ deben pertenecer al segundo.
Por tanto, $H$ debe tener el aspecto\[
H=\left(\begin{array}{cc}
H_{0} & 0\\
0 & H_{1}\end{array}\right)=H_{0}\otimes\Phi+H_{1}\otimes\Phi^{\dagger},\]
mientras que $Q$ y $Q^{\dagger}$ deben ser de la forma\[
\begin{array}{c}
Q=\left(\begin{array}{cc}
0 & Q_{1}\\
Q_{2} & 0\end{array}\right)=Q_{1}\otimes\Psi^{\dagger}+Q_{2}\otimes\Psi\\
\\Q^{\dagger}=\left(\begin{array}{cc}
0 & Q_{1}^{\dagger}\\
Q_{2}^{\dagger} & 0\end{array}\right)=Q_{1}^{\dagger}\otimes\Psi^{\dagger}+Q_{2}^{\dagger}\otimes\Psi\end{array}.\]
La idea central de Witten en \cite{Wit 81} consiste
en tomar la forma más sencilla posible para estos operadores (o supercargas,
en la terminología física). Concretamente, tomaremos\begin{equation}
\begin{array}{c}
Q=\left(\begin{array}{cc}
0 & 0\\
A & 0\end{array}\right)=A\otimes\Psi\\
\\Q^{\dagger}=\left(\begin{array}{cc}
0 & A^{\dagger}\\
0 & 0\end{array}\right)=A^{\dagger}\otimes\Psi^{\dagger}\end{array},\label{eq:QS}\end{equation}
donde $A,A^{\dagger}\in End_{\mathbb{C}}(L^{2}(\mathbb{R}))$ son
operadores que habrá que determinar. Para ello, veamos qué condiciones
imponen las ecuaciones a las que queremos llegar. De $[\![Q^{\dagger},Q]\!]=H$,
resulta:\begin{equation}
\left(\begin{array}{cc}
H_{0} & 0\\
0 & H_{1}\end{array}\right)=H=[\![Q^{\dagger},Q]\!]=Q^{\dagger}\circ Q+Q\circ Q^{\dagger}=\left(\begin{array}{cc}
A^{\dagger}A & 0\\
0 & AA^{\dagger}\end{array}\right).\label{eq:AsyHs}\end{equation}
Vemos entonces que todo resulta sencillo si se cumplen dos condiciones:

\begin{enumerate}
\item Tomamos como componente $H_{0}$ del superhamiltoniano $H$ el propio
operador clásico:\begin{equation}
H_{0}=\widetilde{H}=-\frac{1}{2}\frac{d^{2}}{dx^{2}}+V(x).\label{eq:H0}\end{equation}

\item Se puede encontrar una descomposición de $\widetilde{H}$ en la forma
$A^{\dagger}A$. En ese caso, también conoceremos $H_{1}=AA^{\dagger}$.
\end{enumerate}
El problema de encontrar unos operadores $A,A^{\dagger}$ tales que
$\widetilde{H}=A^{\dagger}A$ es, básicamente, el de hallar una ``raíz
cuadrada'' del Hamiltoniano clásico $\widetilde{H}$. Esto se puede
resolver mediante el procedimiento de ``completar cuadrados'': partimos
de (\ref{eq:H0}); si $V$ \emph{pudiera expresarse} como\begin{equation}
V=-\frac{1}{\sqrt{2}}\frac{dW}{dx}+W^{2},\label{eq:superpotencial}\end{equation}
tendríamos\[
\widetilde{H}=-\frac{1}{2}\frac{d^{2}}{dx^{2}}-\frac{1}{\sqrt{2}}\frac{dW}{dx}+W^{2}\]
y esto, entendiendo el producto de operadores como su composición,
no es otra cosa que\[
\widetilde{H}=\left(-\frac{1}{\sqrt{2}}\frac{d}{dx}+W\right)\cdot\left(\frac{1}{\sqrt{2}}\frac{d}{dx}+W\right).\]
Por tanto, conseguiremos tener $\widetilde{H}=A^{\dagger}A$ si tomamos\[
\begin{array}{c}
A^{\dagger}=-\frac{1}{\sqrt{2}}\frac{d}{dx}+W\\
\\A=\frac{1}{\sqrt{2}}\frac{d}{dx}+W\end{array}\]
siendo $W$ una solución a la ecuación (\ref{eq:superpotencial}).
A $W$ se le denomina en Física el superpotencial.

Volviendo ahora a las ecuaciones (\ref{eq:QS}) y (\ref{eq:AsyHs}),
es fácil darse cuenta%
\footnote{Se deja como ejercicio para el lector comprobar que $AA^{\dagger}=-\frac{1}{2}\frac{d^{2}}{dx^{2}}+\frac{1}{\sqrt{2}}\frac{dW}{dx}+W^{2}$.%
} de que definiendo los operadores $H,Q^{\dagger},Q\in End_{\mathbb{C}}^{G}(\mathcal{H})$
como\[
\begin{array}{c}
H=A^{\dagger}A\otimes\Phi+AA^{\dagger}\otimes\Phi^{\dagger}\\
=(-\frac{1}{2}\frac{d^{2}}{dx^{2}}-\frac{1}{\sqrt{2}}\frac{dW}{dx}+W^{2})\otimes\Phi+(-\frac{1}{2}\frac{d^{2}}{dx^{2}}+\frac{1}{\sqrt{2}}\frac{dW}{dx}+W^{2})\otimes\Phi^{\dagger}\\
\\Q=A\otimes\Psi=(\frac{1}{\sqrt{2}}\frac{d}{dx}+W)\otimes\Psi\\
\\Q^{\dagger}=A^{\dagger}\otimes\Psi^{\dagger}=(-\frac{1}{\sqrt{2}}\frac{d}{dx}+W)\otimes\Psi^{\dagger}\end{array},\]
o bien, en términos matriciales,\begin{equation}
\begin{array}{c}
H=\left(\begin{array}{cc}
A^{\dagger}A & 0\\
0 & AA^{\dagger}\end{array}\right)=\left(\begin{array}{cc}
-\frac{1}{2}\frac{d^{2}}{dx^{2}}-\frac{1}{\sqrt{2}}\frac{dW}{dx}+W^{2} & 0\\
0 & -\frac{1}{2}\frac{d^{2}}{dx^{2}}+\frac{1}{\sqrt{2}}\frac{dW}{dx}+W^{2}\end{array}\right)\\
\\Q=\left(\begin{array}{cc}
0 & 0\\
\frac{1}{\sqrt{2}}\frac{d}{dx}+W & 0\end{array}\right)\\
\\Q^{\dagger}=\left(\begin{array}{cc}
0 & -\frac{1}{\sqrt{2}}\frac{d}{dx}+W\\
0 & 0\end{array}\right)\end{array}\label{eq:HQQ+}\end{equation}
se cumple que\[
\begin{array}{c}
[\![Q^{\dagger},H]\!]=0=[\![H,Q]\!]\\
\\{}[\![Q^{\dagger},Q]\!]=H\end{array},\]
esto es: se tiene una realización explícita del superálgebra de Heisenberg
$\left\langle H,Q^{\dagger},Q\right\rangle $ como subálgebra de $End_{\mathbb{C}}^{G}(\mathcal{H})$.

Finalizaremos esta sección con unas observaciones acerca de otras
convenciones que pueden encontrarse en la literatura sobre SUSY QM.
En Física, además de la base $\widetilde{\mathcal{B}}$ de $End_{\mathbb{C}}^{G}(\mathbb{C}^{1|1})$
que definimos en la sección precedente, también es común utilizar
esta otra:\[
\widetilde{\mathcal{B}}_{\sigma}=\left\{ \sigma_{0}=\left(\begin{array}{cc}
1 & 0\\
0 & 0\end{array}\right),\sigma^{+}=\left(\begin{array}{cc}
0 & 1\\
0 & 0\end{array}\right),\sigma^{-}=\left(\begin{array}{cc}
0 & 0\\
1 & 0\end{array}\right),\sigma_{3}=\left(\begin{array}{cc}
1 & 0\\
0 & -1\end{array}\right)\right\} .\]
Los elementos de esta base, tienen las siguientes propiedades (que
son las que hacen que sean de interés físico):\[
\begin{array}{c}
\sigma^{+}\sigma^{-}+\sigma^{-}\sigma^{+}=[\sigma^{+},\sigma^{-}]_{+}=I=\left(\begin{array}{cc}
1 & 0\\
0 & 1\end{array}\right)\\
\\\sigma^{+}\sigma^{-}-\sigma^{-}\sigma^{+}=[\sigma^{+},\sigma^{-}]\_ =\sigma_{3}=\left(\begin{array}{cc}
1 & 0\\
0 & -1\end{array}\right)\end{array}.\]
Ésta es la base que utilizó Witten en \cite{Wit 81}.
De hecho, él escribió el superhamiltoniano $H$ (omitiendo de nuevo
el factor $\frac{\hbar}{m}$) como\[
H=\frac{1}{2}p^{2}+W^{2}(x)-\frac{1}{\sqrt{2}}\sigma_{3}W'(x),\]
donde el apóstrofe denota derivación. Según lo que hemos visto, esto
debe entenderse como\[
\begin{array}{r}
H=(\frac{1}{2}p^{2}+W^{2}(x))\otimes\left(\begin{array}{cc}
1 & 0\\
0 & 1\end{array}\right)-\frac{1}{\sqrt{2}}W'(x)\otimes\left(\begin{array}{cc}
1 & 0\\
0 & -1\end{array}\right)\\
\\=\left(\begin{array}{cc}
\frac{1}{2}p^{2}+W^{2}(x)-\frac{1}{\sqrt{2}}W'(x) & 0\\
0 & \frac{1}{2}p^{2}+W^{2}(x)+\frac{1}{\sqrt{2}}W'(x)\end{array}\right)\end{array}.\]
Como $p=-i\frac{d}{dx}$ (recuérdese la discusión de la sección \ref{S5}),
resulta:\[
H=\left(\begin{array}{cc}
-\frac{1}{2}\frac{d^{2}}{dx^{2}}+W^{2}(x)-\frac{1}{\sqrt{2}}W'(x) & 0\\
0 & -\frac{1}{2}\frac{d^{2}}{dx^{2}}+W^{2}(x)+\frac{1}{\sqrt{2}}W'(x)\end{array}\right)\]
que es la misma expresión de (\ref{eq:HQQ+}).

\section{Aplicaciones de la supersimetría}

En la sección anterior hemos visto que la clave para construir una
realización de la superálgebra de Heisenberg reside en la introducción
del superpotencial $W.$ A partir de él se definen los operadores
$A$ y $A^{\dagger}$ y, con ellos, las componentes del superhamiltoniano
$H=A^{\dagger}A\otimes\Phi+AA^{\dagger}\otimes\Phi^{\dagger}$. Ahora
bien, desde un punto de vista estrictamente práctico podría objetarse
que este procedimiento tiene poco interés, pues para calcular $W$
hay que resolver la ecuación (\ref{eq:superpotencial}) la cual, a 
primera vista, parece incluso más complicada que la propia
ecuación de Schrödinger con el potencial $V$.

Sin embargo, esto no es así y ahora veremos por qué. Consideremos
el Hamiltoniano clásico inicial,\[
\widetilde{H}_{0}=-\frac{1}{2}\frac{d^{2}}{dx^{2}}+V.\]
Para muchos de los operadores que aparecen en Física, se tiene que
el espectro es discreto y está acotado inferiormente. Suponiendo que
éste es el caso de $\widetilde{H}_{0}$, escribamos su espectro como
$Spec_{\mathbb{R}}(\widetilde{H}_{0})=\{\lambda_{0},\lambda_{1},...,\}$;
si $\lambda_{0}$ es el mínimo de $Spec_{\mathbb{R}}(\widetilde{H}_{0})$,
sustituyendo $\widetilde{H}_{0}$ por $H_{0}=\widetilde{H}_{0}-\lambda_{0}\cdot I$,
el menor valor propio en la ecuación\[
H_{0}\phi_{k}=\lambda_{k}\phi_{k}\]
puede tomarse como $\lambda_{k}=0$. Al vector propio correspondiente,
$\phi_{0}(x)\in L^{2}(\mathbb{R})$, se le denomina estado base. Entonces,
se tiene que $\phi_{0}$ cumple la ecuación\[
\left(-\frac{1}{2}\frac{d^{2}}{dx^{2}}+V\right)\phi_{0}=0,\]
de donde, despejando, el potencial $V$ se puede expresar en términos
del estado base:\[
V(x)=\frac{\phi''_{0}(x)}{2\phi_{0}(x)}.\]

Por otra parte, toda función de onda $\phi(x)$ perteneciente al núcleo
de $A$ cumple que\[
H_{0}\phi=(A^{\dagger}A)\phi=0,\]
es decir, es un elemento del subespacio propio correspondiente al
valor propio $\lambda=0$ de $H_{0}$. Por tanto, si $\phi_{0}\in\ker A$
es $\phi_{0}\in\ker H_{0}$ y automáticamente se cumplen las dos condiciones\begin{equation}
V(x)=\frac{\phi''_{0}(x)}{2\phi_{0}(x)}\label{eq:calculoW}\end{equation}
y\[
W(x)=-\frac{\phi'_{0}(x)}{\sqrt{2}\phi_{0}(x)}=-\frac{1}{\sqrt{2}}\frac{d}{dx}\ln\phi_{0}(x).\]
Esta última expresión%
\footnote{En esta discusión, pasamos por alto detalles importantes como si $W(x)$
presenta singularidades debido a la presencia del logaritmo, etc.
Puede probarse que para la clase de potenciales que se utiliza en
Mecánica Cuántica tales problemas pueden resolverse satisfactoriamente.
Para los detalles remitimos al lector interesado a \cite{CKS 95}.%
}, en particular, muestra que el cálculo del superpotencial $W$ no
es más complicado que el cálculo del subespacio invariante por $H_{0}$
correspondiente a $\lambda=0$.

En lo que sigue, cuando hablemos de estados base de $H_{0}$ consideraremos%
\footnote{Puede probarse que esto no supone ninguna pérdida de generalidad.%
} que $\phi_{0}\in\ker A\subset\ker H_{0}$.

Tenemos entonces el Hamiltoniano inicial $H_{0}$ que puede escribirse
como\[
H_{0}=A^{\dagger}A=-\frac{1}{2}\frac{d^{2}}{dx^{2}}-\frac{1}{\sqrt{2}}\frac{dW}{dx}+W^{2},\]
donde $W$ puede calcularse a partir de (\ref{eq:calculoW}). Recordemos
que también teníamos el Hamiltoniano\[
H_{1}=AA^{\dagger}=-\frac{1}{2}\frac{d^{2}}{dx^{2}}+\frac{1}{\sqrt{2}}\frac{dW}{dx}+W^{2}.\]
Éste, se puede poner en la misma forma que $H_{0}$, como $H_{1}=-\frac{1}{2}\frac{d^{2}}{dx^{2}}+\widetilde{V}$,
si definimos\[
\widetilde{V}=\frac{1}{\sqrt{2}}\frac{dW}{dx}+W^{2}.\]
Los potenciales $V$ y $\widetilde{V}$ se suelen denominar ``compañeros
supersimétricos'' (supersymmetric partners). Fijémonos en lo siguiente:
el superhamiltoniano viene dado, como sabemos, por\[
H=H_{0}\otimes\Phi+H_{1}\otimes\Phi^{\dagger}=\left(\begin{array}{cc}
H_{0} & 0\\
0 & H_{1}\end{array}\right).\]
Los estados impares (o fermiónicos) del espacio de Hilbert $\mathcal{H}=\mathcal{H}_{0}\oplus\mathcal{H}_{1}$
están generados por los autovectores de $H_{1}$, pues $\mathcal{H}_{1}$
es un subespacio invariante por $H_{1}$ (con mayor precisión, invariante
por $H_{1}\otimes\Phi^{\dagger}$), como muestra un sencillo cálculo:\[
(H_{1}\otimes\Phi^{\dagger})\left(\phi\otimes\left(\begin{array}{c}
0\\
1\end{array}\right)\right)=\left(\begin{array}{cc}
0 & 0\\
0 & H_{1}\end{array}\right)\left(\begin{array}{c}
0\\
\phi\end{array}\right)=\left(\begin{array}{c}
0\\
H_{1}\phi\end{array}\right)=H_{1}\phi\otimes\left(\begin{array}{c}
0\\
1\end{array}\right).\]
Análogamente, $\mathcal{H}_{0}$ es un subespacio invariante por $H_{0}$
(i.e, por $H_{0}\otimes\Phi$):\[
(H_{0}\otimes\Phi)\left(\phi\otimes\left(\begin{array}{c}
1\\
0\end{array}\right)\right)=\left(\begin{array}{cc}
H_{0} & 0\\
0 & 0\end{array}\right)\left(\begin{array}{c}
\phi\\
0\end{array}\right)=\left(\begin{array}{c}
H_{0}\phi\\
0\end{array}\right)=H_{0}\phi\otimes\left(\begin{array}{c}
1\\
0\end{array}\right),\]
luego los estados pares (o bosónicos) de $\mathcal{H}$ están generados
por los autovectores de $H_{0}$. Ahora, aquí entra en juego el hecho
de que tenemos una representación de la superálgebra de Heisenberg
sobre $\mathcal{H}=\mathcal{H}_{0}\oplus\mathcal{H}_{1}$: como consecuencia
de la existencia de esta representación los espectros de $H_{0}$
y de $H_{1}$ resultan ser biyectivos. Esto se debe a que las supercargas
$Q$ y $Q^{\dagger}$ mezclan ambos tipos de estados (y, por ello,
se dice que son operadores de \emph{supersimetría}):\begin{equation}
Q^{\dagger}\left(\begin{array}{c}
0\\
\phi\end{array}\right)=\left(\begin{array}{cc}
0 & A^{\dagger}\\
0 & 0\end{array}\right)\left(\begin{array}{c}
0\\
\phi\end{array}\right)=\left(\begin{array}{c}
A^{\dagger}\phi\\
0\end{array}\right)\label{eq:supersym1}\end{equation}
\begin{equation}
Q\left(\begin{array}{c}
\phi\\
0\end{array}\right)=\left(\begin{array}{cc}
0 & 0\\
A & 0\end{array}\right)\left(\begin{array}{c}
\phi\\
0\end{array}\right)=\left(\begin{array}{c}
0\\
A\phi\end{array}\right).\label{eq:supersym2}\end{equation}
Consecuentemente, tenemos una degeneración: los estados de la forma\[
\left(\begin{array}{c}
\phi_{k}^{(0)}\\
0\end{array}\right)\in\mathcal{H}_{0},\]
donde $\phi_{k}^{(0)}$ es una función propia del Hamiltoniano $H_{0}$
con valor propio $\lambda_{k}$, y los de la forma\[
\left(\begin{array}{c}
0\\
A\phi_{k}^{(0)}\end{array}\right)\in\mathcal{H}_{1}\]
 son funciones propias del superhamiltoniano $H=H_{0}\otimes\Phi+H_{1}\otimes\Phi^{\dagger}$
con el mismo valor propio $\lambda_{k}$. En efecto:\[
H\left(\begin{array}{c}
\phi_{k}^{(0)}\\
0\end{array}\right)=\left(\begin{array}{c}
H_{0}\phi_{k}^{(0)}\\
0\end{array}\right)=\left(\begin{array}{c}
\lambda_{k}\phi_{k}^{(0)}\\
0\end{array}\right)=\lambda_{k}\left(\begin{array}{c}
\phi_{k}^{(0)}\\
0\end{array}\right)\]
y
\[
H\left(\begin{array}{c}
0\\
A\phi_{k}^{(0)}\end{array}\right)=\left(\begin{array}{c}
0\\
H_{1}A\phi_{k}^{(0)}\end{array}\right)=\left(\begin{array}{c}
0\\
AH_{0}\phi_{k}^{(0)}\end{array}\right)=\lambda_{k}\left(\begin{array}{c}
0\\
A\phi_{k}^{(0)}\end{array}\right).
\]

Insistimos en que\[
\left(\begin{array}{c}
\phi_{k}^{(0)}\\
0\end{array}\right)\; y\;\left(\begin{array}{c}
0\\
A\phi_{k}^{(0)}\end{array}\right)\]
son de distinta paridad, aun cuando son funciones propias del superhamiltoniano
con el mismo valor propio, y que los operadores $Q^{\dagger}$ y $Q$
(de acuerdo con (\ref{eq:supersym1}), (\ref{eq:supersym2})) actúan
como supersimetrías, transformando estos estados entre ellos.

Para recapitular, tenemos que dada una función propia de $H_{0}$
con valor propio $\lambda_{k}$ ($k>0$), $\phi_{k}^{(0)}$, la función
$A\phi_{k}^{(0)}$ es propia de $H_{1}$ con el mismo valor propio%
\footnote{Otra forma de ver esto es mediante el cálculo\[
H_{1}A\phi_{k}^{(0)}=(AA^{\dagger})A\phi_{k}^{(0)}=AH_{0}\phi_{k}^{(0)}=A(\lambda_{k}\phi_{k}^{(0)})=\lambda_{k}A\phi_{k}^{(0)}.\]
}. Con un cálculo análogo, que se deja a cargo del lector, se prueba
que podemos recuperar las funciones propias de $H_{0}$, excepto el
estado base $\phi_{0}^{(0)}$, a partir de las de $H_{1}$: si $\phi_{k}^{(1)}$
($k>0$) es una función propia de $H_{1}$ con valor propio $\lambda_{k}$,
entonces $A^{\dagger}\phi_{k}^{(1)}$ es una función propia de $H_{0}$
con el mismo valor propio.

Los espectros de $H_{0}$ y $H_{1}$, por tanto, son biyectivos. La
única diferencia entre ellos es que el de $H_{1}$ está ``desplazado''
con respecto al de $H_{0}.$ En efecto, el estado base de $H_{1}$
es $A\phi_{1}^{(0)}$ y no $A\phi_{0}^{(0)}$, pues recordemos que
$\phi_{0}^{(0)}\in\ker A\subset\ker H_{0}$, luego $A\phi_{0}^{(0)}=0$
y este estado no puede tomarse como propio de ningún operador.

Así pues, la existencia de supersimetrías en Mecánica Cuántica lleva
aparejada la existencia de funciones propias degeneradas del operador
superhamiltoniano $H$. Veamos cómo puede aprovecharse esto en la
práctica.

Consideremos el potencial de pozo cuadrado infinito que ya estudiamos
en la sección \ref{S8}:
\[
V(x)=\left\{ 
\begin{array}{l}
\infty\text{ si }x\geq L\text{ \'{o} }x\leq0\\
\\
0\text{ si }0<x<L,
\end{array}\right.
\]
y el Hamiltoniano
\[
\widetilde{H}_{0}=-\frac{1}{2}\frac{d^{2}}{dx^{2}}+V(x)
\]
 a que da lugar. Ya vimos que la función de onda correspondiente al
estado base estacionario (no nos interesará ahora la evolución temporal),
haciendo $\hbar=1=m$, es la que corresponde al valor $n=1$:\[
\phi_{1}(x)=\sqrt{\frac{2}{L}}\sin\left(\frac{\pi x}{L}\right),\;0\leq x\leq L.\]
El valor de la energía asociado a este estado es\[
E_{1}=\frac{\pi^{2}}{2L^{2}}.\]
De acuerdo con lo que hemos establecido al principio de esta sección,
le restamos a $\widetilde{H}_{0}$ esta energía del estado base y
obtenemos el nuevo Hamiltoniano
\[
H_{0}=\widetilde{H}_{0}-\frac{\pi^{2}}{2L^{2}}\cdot I,
\]
para el cual las funciones propias son
\begin{equation}
\phi_{k}^{(0)}(x)=\sqrt{\frac{2}{L}}\sin\left(\frac{(k+1)\pi x}{L}\right),\;0\leq x\leq L,k\geq0,\label{eq:fdo}
\end{equation}
y los valores propios correspondientes:\[
\lambda_{k}=E_{k}=\frac{k(k+2)\pi^{2}}{2L^{2}}.\]
El superpotencial $W(x)$ se calcula fácilmente a partir de la expresión
\eqref{eq:superpotencial}:
\[
W(x)=-\frac{1}{\sqrt{2}\phi_{0}^{(0)}(x)}\frac{d}{dx}\phi_{0}^{(0)}(x)=-\frac{\pi}{L\sqrt{2}}\cot\frac{\pi x}{L}.\]
El compañero supersimétrico del potencial pozo cuadrado infinito resulta
ser, entonces,\[
\widetilde{V}(x)=\frac{1}{\sqrt{2}}\frac{dW}{dx}+W^{2}=\frac{\pi^{2}}{2L^{2}}\left(2\csc\left(\frac{\pi x}{L}\right)-1\right).\]
De acuerdo con lo que hemos visto, el Hamiltoniano\[
H_{1}=-\frac{1}{2}\frac{d^{2}}{dx^{2}}+\widetilde{V}(x)=-\frac{1}{2}\frac{d^{2}}{dx^{2}}+\frac{\pi^{2}}{2L^{2}}\left(2\csc\left(\frac{\pi x}{L}\right)-1\right)\]
tiene el mismo espectro (salvo la energía del estado base) que el
Hamiltoniano original $H_{0}$, mucho más sencillo. Además, las funciones
de onda propias de $H_{1}$ se obtienen a partir de las de $H_{0}$
(véase \eqref{eq:fdo}) aplicando $A$. En particular, las primeras funciones
propias de $H_{1}$ son de la forma (omitiendo factores constantes):\[
\phi_{0}^{(1)}(x)\sim\sin^{2}\left(\frac{\pi x}{L}\right)\]
y\[
\phi_{1}^{(1)}(x)\sim\sin\left(\frac{\pi x}{L}\right)\sin\left(\frac{2\pi x}{L}\right).\]

Para comprender la potencia de este método conviene pensar en el proceso
inverso y suponer que lo que nos dan en principio es la ecuación de
Schrödinger con el Hamiltoniano\[
H_{1}=-\frac{1}{2}\frac{d^{2}}{dx^{2}}+\frac{\pi^{2}}{2L^{2}}\left(2\csc\left(\frac{\pi x}{L}\right)-1\right).\]
\emph{A priori}, pensaríamos que la solución será extremadamente complicada
y un intento de resolver este problema mediante el método clásico
nos convencerá enseguida de que esta suposición es correcta. Sin embargo,
la existencia de una supersimetría subyacente al problema permite
hallar las soluciones de este problema a partir de las de uno muchísimo
más sencillo: el del potencial de pozo cuadrado infinito. De hecho,
\emph{todos} los potenciales resolubles exactamente en Mecánica Cuántica
conocidos hasta el momento pueden calcularse mediante el método supersimétrico. La idea consiste
en iterar el proceso: de $H_{0}$ por factorización obtenemos $H_{1}$. Ahora, tomamos a éste
como punto de partida y factorizándolo obtenemos otro Hamiltoniano $H_{2}$, etc. Los
espectros de la cadena $H_{0},H_{1},H_{2}$... están relacionados entre sí, y cada uno
tiene un estado propio menos que el anterior. Si el Hamiltoniano inicial tenía un
número finito de estados propios, este método (llamado de la jerarquía) proporciona una forma
algorítmica de calcularlos todos de una manera sencilla.\\
Para ampliar información sobre este tema y otras aplicaciones de la
supersimetría que no mencionaremos aquí, remitimos al lector a \cite{CKS
95} y la versi\'on ampliada \cite{CKS 01}. Las extensiones del m\'etodo de factorizaci\'on descrito en estas notas
(t\'ecnica de Mielnik), pueden consultarse en \cite{Fer 09}.

\section*{Apéndice A. Las relaciones de incertidumbre\label{apendiceA}}

Supongamos dos observables $A,B\in End_{\mathbb{C}}(\mathcal{H})$ (que, recordemos, son autoadjuntos) 
tales que su conmutador viene dado por $[A,B]\_ =c\cdot I$ con $c \in \mathbb{C}$ e $I$ la identidad.
Dado un estado (esto es, una función de onda normalizada) $\Psi \in \mathcal{H}$, 
se define la varianza de un operador \emph{cualquiera} $X\in End_{\mathbb{C}}(\mathcal{H})$ en
ese estado, $(\Delta X)_{\Psi}$, como
$$
(\Delta X)_{\Psi}^{2}=\langle X^{2} \rangle_{\Psi} - \langle X \rangle_{\Psi}^{2},
$$
donde, también para un operador arbitrario, $\langle X \rangle_{\Psi}$ es su valor
esperado en el estado $\Psi$, definido por:
$$
\langle X \rangle_{\Psi}=\langle \Psi | X | \Psi \rangle.
$$
Consideremos el elemento $(A+\lambda iB)|\Psi \rangle \in \mathcal{H}$, 
($ \lambda \in \mathbb{R}$) cuya norma respecto
del producto escalar en $\mathcal{H}$ (dada por $\langle(A+ \lambda iB)^{2}\rangle_{\Psi} $)
debe ser mayor o igual que cero. Desarrollando esta condición, se tiene:
$$
0 \leq \langle A^{2} + \lambda^{2}B^{2} +i\lambda C \rangle_{\Psi}=
\langle A^{2}\rangle_{\Psi} + \lambda^{2}\langle B^{2}\rangle_{\Psi} +ic\lambda .
$$
Esto significa que el discriminante de esta ecuación ha de ser mayor o igual que
cero, lo cual implica que:
$$
\langle A^{2}\rangle_{\Psi} \langle B^{2}\rangle_{\Psi} \geq \frac{|c|^{2}}{4}.
$$
Definiendo los nuevos operadores
\begin{eqnarray*}
\hat{A}=A-\langle A \rangle_{\Psi} \\
\hat{B}=B-\langle B \rangle_{\Psi}
\end{eqnarray*}
la propiedad $[ \hat{A},\hat{B} ]\_ =c\cdot I$ se sigue cumpliendo, pues $A$ y $B$ sólo 
se han modificado en un factor constante (que conmuta con cualquier operador).
Así, también se cumplirá
$$
\langle \hat{A}^{2}\rangle_{\Psi} \langle \hat{B}^{2}\rangle_{\Psi} \geq \frac{|c|^{2}}{4}.
$$
Pero, desarrollando $\langle \hat{A}^{2}\rangle_{\Psi}$ y $\langle \hat{B}^{2}\rangle_{\Psi}$
de acuerdo con sus definiciones, se obtiene (como es inmediato comprobar):
\begin{eqnarray*}
\langle \hat{A}^{2}\rangle_{\Psi}=\langle A^{2} \rangle_{\Psi} - \langle A \rangle_{\Psi}^{2}=(\Delta A)_{\Psi}^{2} \\
\langle \hat{B}^{2}\rangle_{\Psi}=\langle B^{2} \rangle_{\Psi} - \langle B \rangle_{\Psi}^{2}=(\Delta B)_{\Psi}^{2},
\end{eqnarray*}
de donde se sigue que
$$
(\Delta A)_{\Psi} \cdot (\Delta B)_{\Psi} \geq \frac{|c|}{2},
$$
que es el enunciado matemático de las \emph{relaciones de incertidumbre}
\footnote{En Iberoamérica es frecuente hablar de las ``relaciones de incerteza''.
Ambas formas son válidas, si bien la Real Academia de la Lengua dice que ``incerteza''
es una forma ``en femenino poco usada'' de ``incertidumbre''.} de Heisenberg.\\
La denominación se debe a que $(\Delta A)_{\Psi}$ se interpreta como la incertidumbre 
en la medida de la magnitud representada por el observable $A$ en el estado $\Psi$.
Si aplicamos el cálculo anterior a los operadores posición y momento, que cumplen
$$
[x,-i\hbar\frac{d}{dx}]\_ =i\hbar \cdot I,
$$
resulta que, en cualquier estado $\Psi \in \mathcal{H}$,
$$
(\Delta x)_{\Psi} \cdot (\Delta p)_{\Psi} \geq \frac{\hbar}{2}.
$$
La interpretación física de estas desigualdades es que no es posible determinar con
precisión arbitraria simultáneamente la posición y el momento (equivalentemente, 
la velocidad) de una partícula cuántica. Como consecuencia, no
es posible hablar de la trayectoria definida de una partícula si a la vez se espera
trabajar con su velocidad o momento, como ocurre en el caso de la Dinámica Cuántica
(aunque no hay obstáculos para considerar la trayectoria de una partícula de la
cual se renuncia a conocer su velocidad).

\section*{Apéndice B. El corchete de Poisson y el conmutador cuántico\label{apendiceB}}

Una herramienta básica en la formulación de la Mecánica Clásica, como
hemos visto, es el corchete de Poisson $\{,.,\}$. Si $M$ es el espacio
fásico correspodiente a un cierto sistema físico, un observable físico
es cualquier función $f\in\mathcal{C}^{\infty}(M)$. Desde un punto
de vista algebraico, el corchete de Poisson convierte al anillo de
observables (cuyo producto es el producto de funciones reales) en
un álgebra de Lie $(\mathcal{C}^{\infty}(M),\{.,.\})$, es decir,
se tienen las propiedades (para todas $f$,$g$,$h$$\in\mathcal{C}^{\infty}(M)$):
\begin{enumerate}
\item $\{.,.\}$ es una aplicación $\mathbb{R}-$bilineal antisimétrica.
\item Se cumple la llamada regla de Leibniz:\[
\{f,g\cdot h\}=g\cdot\{f,h\}+\{f,g\}\cdot h.\]

\item Se cumple la identidad de Jacobi:\[
\{f,\{g,h\}\}=\{\{f,g\},h\}+\{g,\{f,h\}\}.\]
\end{enumerate}
Dirac, en su conocido libro sobre los principios de la Mecánica Cuántica
(véase \cite{Dir 81}), se planteó cuál sería el análogo cuántico de esta estructura algebraica
presente en el caso clásico. Ahora bien, como ya se ha mencionado,
los observables son operadores $A\in End_{\mathbb{C}}(\mathcal{H})$ sobre un cierto
espacio de Hilbert complejo, que también forman un anillo (con respecto
a la composición), solo que éste es no conmutativo. La idea de Dirac
consistió en buscar un nuevo corchete\[
[.,.]:End_{\mathbb{C}}(\mathcal{H})\times End_{\mathbb{C}}(\mathcal{H})\rightarrow End_{\mathbb{C}}(\mathcal{H})\]
que mantenga las propiedades del de Poisson, esto es, que convierta
$(End_{\mathbb{C}}(\mathcal{H}),[.,.])$ en un álgebra de Lie. Para construirlo,
Dirac tuvo en cuenta la no conmutatividad del anillo $End_{\mathbb{C}}(\mathcal{H})$
y calculó (para $A_{i},B_{i}\in End_{\mathbb{C}}(\mathcal{H})$, $i\in\{1,2\}$)
el corchete $[A_{1}A_{2},B_{1}B_{2}]$ de dos formas distintas, aplicando
reiteradamente la regla de Leibniz%
\footnote{Por simplicidad en la escritura, indicaremos la composición de dos
operadores mediante la yuxtaposición, es decir, $AB$ denotará $A\circ B$.%
}:
\[
\begin{array}{ccl}
& [A_{1}A_{2},B_{1}B_{2}] = \\
& [A_{1},B_{1}B_{2}]A_{2}+A_{1}[A_{2},B_{1}B_{2}] = \\
& [A_{1},B_{1}]B_{2}A_{2}+B_{1}[A_{1},B_{2}]A_{2}+A_{1}[A_{2},B_{1}]B_{2}+A_{1}B_{1}[A_{2},B_{2}]
\end{array}
\]
y por otra parte:
\[
\begin{array}{ccl}
& [A_{1}A_{2},B_{1}B_{2}] = \\
& [A_{1}A_{2},B_{1}]B_{2}+B_{1}[A_{1}A_{2},B_{2}] = \\
& [A_{1},B_{1}]A_{2}B_{2}+A_{1}[A_{2},B_{1}]B_{2}+B_{1}[A_{1},B_{2}]A_{2}+B_{1}A_{1}[A_{2},B_{2}].
\end{array}
\]
Igualando ambas expresiones, se obtiene\[
[A_{1},B_{1}](A_{2}B_{2}-B_{2}A_{2})=(A_{1}B_{1}-B_{1}A_{1})[A_{2},B_{2}].\]
Pero al ser $A_{1},A_{2},B_{1},B_{2}$ independientes entre sí, esta
relación implica\footnote{En realidad, la afirmaci\'on de Dirac necesita justificaci\'on, ya que no es cierta en \'algebras de Poisson arbitrarias. Lo que ocurre es que se est\'a haciendo la suposici\'on impl\'icita de que existen unos elementos particulares $Q,P\in \mathrm{Obs}(\mathcal{H})$ tales que $[P,Q]=u$, donde 
$u$ es un elemento unidad del \'algebra $\mathrm{Obs}(\mathcal{H})$. A partir de aqu\'i, se puede justificar la afirmaci\'on de Dirac.} que para cualesquiera $A,B\in End_{\mathbb{C}}(\mathcal{H})$,\[
[A,B]=k(AB-BA)\]
para una cierta constante $k\in\mathbb{C}$. Es inmediato que tomando
esta definición para $[.,.]$, se cumplen las propiedades de bilinealidad,
antisimetría, Leibniz y Jacobi. Faltaría determinar la constante $k$.

Observemos que una condición adicional razonable es que si $A$ y
$B$ son operadores autoadjuntos, $[A,B]$ sea autoadjunto también
(pues, desde un punto de vista físico, $[A,B]$ proporciona el resultado
de realizar las medidas correspondientes a $A$ y $B$ consecutivamente).
Pero se tiene que:\[
\begin{array}{rllll}
[A,B]^{\dagger} & = & k^{*}((AB)^{\dagger}-(BA)^{\dagger}) & = & k^{*}(B^{\dagger}A^{\dagger}-A^{\dagger}B^{\dagger})\\
 & = & k^{*}(BA-AB) & = & -k^{*}(AB-BA)\end{array}\]
luego debe ser\[
k=\frac{i}{\hbar},\]
para una cierta constante real $\hbar\in\mathbb{R}$ (experimentalmente
se determina \emph{a posteriori} que esta constante es, en efecto,
la constante de Planck).

Así pues, Dirac introdujo el álgebra de Lie $(End_{\mathbb{C}}(\mathcal{H}),[.,.])$ de los 
observables cuánticos. En Física, al corchete
\[
[A,B]=\frac{i}{\hbar}(A\circ B-B\circ A)
\]
se le denomina \emph{conmutador canónico o de Dirac}.


\begin{thebibliography}{}
\bibitem[Azc-Izq 98]{Azc-Izq 98} J. A. de Azcárraga, J. M. Izquierdo: \emph{Lie Groups, Lie Algebras, 
Cohomology and Some Applications in Physics}. Cambridge Monographs in Mathematical Physics, CUP (1998).

\bibitem[Ber 66]{Ber 66} F. A. Berezin: \emph{The method of second
quantization}. Academic Press (1966).

\bibitem[Ber 87]{Ber 87} F. A. Berezin: \emph{Introduction to Superanalysis}.
Reidel Publ. Dordretch (1987).

\bibitem[Ber-Mar 75]{Ber-Mar 75} F. A. Berezin and M. S. Marinov: 
\emph{Classical spin and Grassmann algebra}. JETP Letters \textbf{21} n11 (1975) 320-322.

\bibitem[BDZVH 76]{BDZVH 76} L. Brink, S. Deser, P. Di Vecchia, P.
Van Hove and B. Zumino: \emph{Local supersymmetry for spinning particles}. 
Phys. Lett. \textbf{64B} (1976) 435.

\bibitem[Cas 76]{Cas 76} R. Casalbuoni:
\emph{On the quatization of Systems with
anticommuting variables}. Nuovo Cim. \textbf{33} (1976) 115-124.

\bibitem[CKS 95]{CKS 95} F. Cooper, A. Khare and U. Sukhatme:
\emph{Supersymmetry and Quantum Mechanics}. Phys. Rep. \textbf{251}
(1995) 267-385.

\bibitem[CKS 01]{CKS 01}F. Cooper, A. Khare and U. Sukhatme:
\emph{Supersymmetry in quantum mechanics}. World Scientific (2001).

\bibitem[CNS 75]{CNS 75} L. Corwin, Y. Ne'eman and S. Sternberg:
\emph{Graded Lie algebras in mathematics and physics (Bose-Fermi symmetry)}.
Rev. Mod. Phys. \textbf{47} (1975) 573-604.

\bibitem[Cro-Rit 83]{Cro-Rit 83} M. de Crombrugghe and V. Rittenberg:
\emph{Supersymmetric quantum mechanics}. Ann. of Phys. \textbf{151} (1983) 99-126.

\bibitem[Dir 81]{Dir 81} P. A. M. Dirac: \emph{Principles of Quantum Mechanics}.
Oxford UP (1981).

\bibitem[Fer 09]{Fer 09} D.J. Fernandez:
\emph{Supersymmetric Quantum Mechanics}.
CINVESTAV Advanced Summer School 2009, Mexico City, Mexico.
arXiv:0910.0192

\bibitem[Fre 86]{Fre 86} P.G.O. Freund:
\emph{Introduction to supersymmetry}. Cambridge Univ. Press (1986).

\bibitem[FSS 00]{FSS 00} L. Frappat, A. Sciarrino and P. Sorba:
\emph{Dictionary on Lie algebras and superalgebras}.
Academic Press (2000).

\bibitem[Gas 03]{Gas 03} S. Gasiorowicz: \emph{Quantum Physics},
3rd. Edition. John Wiley \& Sons. New York (2003).

\bibitem[Ger-Ste 24]{Ger-Ste 24} W. Gerlach and O. Stern: \emph{Das magnetische 
Moment des Silberatoms}. Zeitschrift für Physik \textbf{9} (1922) 353-355, and
\emph{The directional quantisation in the magnetic field}. Ann. Phys. \textbf{74} 
(1924) 673-697.

\bibitem[Gen-Kri 85]{Gen-Kri 85} L.E. Gendenshtein and I.V. Krive:
\emph{Supersymmetry in quantum mechanics}. Soviet Phys. Uspekhi \textbf{28} (1985) 645-666.

\bibitem[Gou-Uhl 25]{Gou-Uhl 25} S. Goudsmith and G. E. Uhlenbeck:
\emph{Ersetzung der Hypothese vom unmechanischen Zwang durch eine Forderung 
bezüglich des inneren Verhaltens jedes einzelnen Elektrons}. Naturwiss. \textbf{XIII} (1925) 953, 
and \emph{Spinning Electrons and the Structure of Spectra}. Nature \textbf{CXVII} (1926) 264.

\bibitem[GSW 87]{GSW 87} M.B. Green, J.H. Schwarz and E. Witten:
\emph{Superstring Theory}, in two volumes. Cambridge University Press
(1987).

\bibitem[Kib-Dao 04]{Kib-Dao 04} M. Kibler and M. Daoud: \emph{On supersymmetric Quantum Mechanics}.
arXiv: quant-ph/049169.

\bibitem[Kos 77]{Kos 77} B. Kostant: Differential geometrical methods
in Math. Phys. (Proc. Sympos, Univ. Bonn, 1975) 177. Lecture Notes
in Math, Vol \textbf{570}, Springer, Berlin (1977).

\bibitem[Sche 79]{Sche 79} M. Scheunert: \emph{The theory of Lie superalgebras}.
Lecture Notes in Math, Vol \textbf{716}, Springer, Berlin (1979)

\bibitem[Str-Wig 89]{Str-Wig 89} R. F. Streater and A. S. Wightman:
\emph{PCT, Spin and Statistics, and All That}. Addison-Wesley, New
York (1989).

\bibitem[Von 32]{Von 32} J. Von Neumann:
\emph{Fundamentos Matemáticos de la Mecánica Cuántica}. CSIC, Madrid (1991).

\bibitem[Wes-Bag 92]{Wes-Bag 92} J. Wess and J. Bagger: \emph{Supersymmetry
and Supergravity}, 2nd edition. Princeton University Press (1992).

\bibitem[Wit 92]{Wit 92} B. de Witt: \emph{Supermanifolds}. Cambridge Univ. Press 2nd Edition (1992).

\bibitem[Wit 81]{Wit 81} E. Witten: \emph{Dynamical breaking of supersymmetry}.
Nucl. Phys. B \textbf{185} (1981) 513-554. 

\bibitem[Yan-Wu 75]{Yan-Wu 75} T.T. Wu and C.N. Yang: \emph{Concept of nonintegrable phase factors and global formulation of gauge fields}. Phys. Rev. \textbf{D12} (1975) 3485-3857.
\end{thebibliography}
\end{document}